\documentclass[12pt]{article}

\usepackage{graphics}
\usepackage{epsfig}

\newcommand{\nn}{\nonumber}

\newcommand{\bq}{\begin{eqnarray} }
\newcommand{\eq}{\end{eqnarray} }
\newcommand{\e}{\epsilon}

\begin{document}\begin{titlepage}

\begin{flushright}
OSU-HEP-03-10\\
August 2003\\
\end{flushright}

\vspace*{2.0cm}
\begin{center}
{\Large {\bf Anomalous $U(1)$ Symmetry and \\Lepton Flavor Violation\\[-0.05in]
} }

\vspace*{2cm}
 {\large {\bf K.S. Babu,\footnote{E-mail address: babu@okstate.edu}
 Ts. Enkhbat\footnote{E-mail address: enkhbat@okstate.edu}
 and I. Gogoladze\footnote{On a leave of absence from:
Andronikashvili Institute of Physics, GAS, 380077 Tbilisi,
Georgia.  \\ E-mail address: ilia@hep.phy.okstate.edu}}}

 \vspace*{1cm}
{\it Department of Physics, Oklahoma State University\\
Stillwater, OK~74078, USA }
\end{center}

 \vspace*{2.0cm}

\begin{abstract}

We show that in a large class of models based on anomalous $U(1)$
symmetry which addresses the fermion mass hierarchy problem,
leptonic flavor changing processes are induced that are in the
experimentally interesting range. The flavor violation occurs
through the renormalization group evolution of the soft SUSY
breaking parameters between the string scale and the $U(1)_A$
breaking scale. We derive general expressions for the evolution of
these parameters in the presence of higher dimensional operators.
Several sources for the flavor violation are identified:
flavor--dependent contributions to the soft masses from the
$U(1)_A$ gaugino, scalar mass corrections proportional to the
trace of $U(1)_A$ charge, non--proportional $A$--terms from vertex
corrections, and the $U(1)_A$ $D$--term. Quantitative estimates
for the decays $\mu\rightarrow e\gamma$ and $\tau\rightarrow \mu
\gamma$ are presented in supergravity models which accommodate the
relic abundance of neutralino dark matter.

\end{abstract}

\end{titlepage}

\newpage

\section{Introduction}

The Standard Model, while highly successful in explaining all
experimental data, does not provide an explanation for the
observed hierarchy in the masses and mixings of quarks and
leptons. Extended symmetries are often speculated to address this
problem. Family--dependent $U(1)$ symmetry is a widely studied
extension. An attractive scenario is the Froggatt--Nielsen scheme
\cite{FrNl}. In this scenario all the Yukawa couplings are assumed
to be of order one, but the ones which generate the light fermion
masses arise only as nonrenormalizable operators suppressed by
powers of a small parameter $\e\equiv\langle S\rangle/M$, where
$\langle S\rangle$ is the flavor symmetry breaking order parameter
and $M$ is a more fundamental mass scale. With the flavor $U(1)$
charges of fermions differing only by order one, large hierarchy
factors, such as $m_u/m_t\sim10^{-6}$, are explained.

A natural origin for the flavor $U(1)$ symmetry is the anomalous
$U(1)_A$ gauge symmetry of string theory \cite{GS}. The small
expansion parameter $\e$ arises in a natural way in anomalous
$U(1)$ models through the Fayet--Iliopoulos term induced by the
gravitational anomaly \cite{DSW}. Such models have been
extensively studied in the literature for understanding the
fermion mass hierarchy puzzle \cite{lfU1,bgw}. The purpose of this
paper is to address flavor changing neutral currents in this class
of models.

In the presence of low energy supersymmetry (SUSY), a
family--dependent anomalous $U(1)_A$ symmetry will induce flavor
changing processes, even when there is no flavor violation in the
soft SUSY breaking parameters at the fundamental scale -- taken to
be the string scale. Such violations will be generated through the
renormalization group evolution (RGE) of the SUSY breaking
parameters between $M_{string}$ and the $U(1)_A$ breaking scale
$\langle S\rangle$. We derive general expressions for the
evolution of these parameters in the presence of higher
dimensional operators. Our results  can be applied to a wide class
of Froggatt--Nielsen models. We have found several sources of
flavor violation. In the momentum interval $\langle S\rangle\leq
\mu \leq M_{string}$,  the $U(1)_A$ gaugino is active and will
contribute differently to the soft masses of different families.
Because ${\rm Tr}U(1)_A$ is not zero in anomalous $U(1)$ models,
there are nonuniversal RGE contributions to the soft scalar masses
arising from the $D$--term proportional to the respective flavor
charges. Furthermore, the trilinear $A$--terms will receive vertex
corrections from the $U(1)_A$ gaugino that are not proportional to
the respective Yukawa couplings. In addition to these RGE effects,
upon symmetry breaking, the $D$--term associated with the $U(1)_A$
will also induce nonuniversal masses for the sfermions
\cite{BabuMohapatra1}--\cite{WellsTobe2}.

In this paper we investigate the combined effects of
nonuniversality for lepton flavor violating (LFV) decays
$\mu\rightarrow e\gamma$ and $\tau\rightarrow \mu \gamma$ in a
class of anomalous $U(1)_A$ models. Quantitative predictions for
the branching ratios are presented in two specific models of
fermion mass hierarchy. We find that the branching ratio for
$\mu\rightarrow e\gamma$ is around the current experimental limit,
while $\tau\rightarrow \mu \gamma$ may be accessible in the
future. In our analysis we also include the right--handed
neutrino--induced LFV effects, which have been widely studied in
the literature \cite{Hisano}--\cite{babudutta}. These effects turn
out to be significant in some but not all cases that we study. In
fact, within our framework, if leptogenesis is assumed to be the
source of the observed cosmological baryon asymmetry \cite{FY}, it
turns out that the right--handed neutrino induced LFV effects are
negligible. Related effects from SUSY GUT thresholds for LFV have
been studied in Ref. \cite{BarHall}. The flavor violating effects
are more prominent in the leptonic sector compared to the quark
sector since quark flavor violation is diluted somewhat by the
gluino focusing effects which arise during the evolution of the
SUSY breaking parameters below the $U(1)_A$ breaking scale. This
is especially so when the SUSY parameter are chosen, as we do,
such that the lightest neutralino is the dark matter with an
acceptable cosmological abundance.

The structure of the paper is as follows. In Section 2 we
introduce our models. In 2.1 we describe our models of fermion
mass hierarchy, in 2.2 we present the details of  anomalous $U(1)$
models. In 2.3 we present our fermion mass fits for two models,
Model 1 and Model 2. In 2.4 we discuss baryogenesis through
leptogenesis in these models. In Section 3 we derive generalized
expressions for the evolution of the soft SUSY breaking parameters
appropriate for a class of Froggatt--Nielsen models. Section 4 is
devoted to the numerical analysis of the branching ratios for the
processes $\mu\rightarrow e\gamma$ and $\tau\rightarrow\mu\gamma$.
In 4.1 we outline the qualitative features of flavor violation
arising from various sources. 4.2 has our numerical fits. Figures
3 and 4 show our results for the branching ratios for
$\mu\rightarrow e\gamma$ in the two models of fermion masses. In
4.3 we address the $D$-term splitting problem. Our conclusions are
given in Section 5.

\section{Fermion Mass Matrices and Anomalous $U(1)_A$ Flavor Symmetry }

In this section we present two specific models of fermion mass
hierarchy based on a flavor--dependent anomalous $U(1)_A$ symmetry
using the Froggatt--Nielsen mechanism \cite{FrNl}. The $U(1)_A$
symmetry is broken by an MSSM singlet flavon filed $S$ slightly
below the string scale ($M_{st}$). This provides the small
expansion parameter $\e = \langle S\rangle/M_{st}$ needed for
explaining the fermion mass hierarchy. We also present  explicit
models of anomalous $U(1)_A$ symmetry.

\subsection{Fermion Mass Hierarchy}

A general form of the superpotential which can explain the fermion
masses and mixing hierarchy through the Froggatt--Nielsen
mechanism has the form
\begin{eqnarray}\label{superP1}
W&=&\frac{y_{ij}^u}{n^u_{ij}!} Q_i u_j^c H_u
\left(\frac{S}{M_{st}}\right)^{n^u_{ij}}+
\frac{y_{ij}^d}{n^d_{ij}!}
Q_id_j^c H_d \left(\frac{S}{M_{st}}\right)^{n^d_{ij}}\nn \\
&+& \frac{y_{ij}^e}{n^e_{ij}!} L_i e_j^c H_d
\left(\frac{S}{M_{st}}\right)^{n^e_{ij}}
 + \frac{y_{ij}^\nu}{n^\nu_{ij}!} L_i \nu^c_j H_u \left(\frac{S}{M_{st}}\right)^{n^\nu_{ij}}\nonumber\\
 &+&{M_{R}}_{ij} \nu^c_i \nu^c_j\left(\frac{S}{M_{st}}\right)^{n^{\nu^c}_{ij}}+\mu H_u
 H_d\, ,
\end{eqnarray}
where $i,j=(1,\,2,\,3)$ are family indices, ${n^u_{ij}}$,
${n^d_{ij}}$, ${n^e_{ij}}$, ${n^{\nu}_{ij}}$ and $n^{\nu^c}_{ij}$
are positive integers fixed by the choice of $U(1)_A$ charge
assignment. $y^u_{ij}$ etc. are Yukawa coupling coefficients which
are all taken to be of order one. Here $M_R$ is the right-handed
($\nu^c_i$) neutrino mass scale. We use the standard notation for
the MSSM fields.

The soft supersymmetry breaking terms which will induce LFV have
the form given by
\begin{eqnarray}\label{lagr1}
-\textit{L}_{soft}&=&\left\{\frac{a_{ij}^u}{n^u_{ij}!} \tilde{Q}_i
\tilde{u}_j^c H_u \left(\frac{S}{M_{st}}\right)^{n^u_{ij}}+
\frac{a_{ij}^d}{n^d_{ij}!}
\tilde{Q}_i\tilde{d}_j^c H_d \left(\frac{S}{M_{st}}\right)^{n^d_{ij}}\nn \right.\\
&+& \frac{a_{ij}^e}{n^e_{ij}!} \tilde{L}_i \tilde{e}_j^c H_d
\left(\frac{S}{M_{st}}\right)^{n^e_{ij}}
 + \frac{a_{ij}^\nu}{n^\nu_{ij}!} \tilde{L}_i {\tilde{\nu}}^c_{j} H_u
 \left(\frac{S}{M_{st}}\right)^{n^\nu_{ij}}+B^\prime {M_R}_{ij}\tilde{\nu}^c_i\tilde{\nu}^c_j\left(\frac{S}{M_{st}}\right)^{n^{\nu^c}_{ij}}\nn\\
 &+&\left.\frac{1}{2}\sum_{i=1,2,3}M^i_{1/2}\lambda_i\lambda_i+\frac{1}{2}M_{\lambda_F}\lambda_F\lambda_F+B\mu H_u
 H_d+\mbox{h. c.}\right\}
 \nn\\
 &+&\sum_{}\left(\tilde{m}^2_f\right)_{ab}
 \tilde{f}_a^\dagger\tilde{f}_b
 +\tilde{m_s}^2|S|^2+\tilde{m}^2_{H_{u}}|H_u|^2+\tilde{m}^2_{H_{d}}|H_d|^2\,.
\end{eqnarray}
Here a tilde stands for the scalar components of the matter
superfields, and $\lambda_i$ and $\lambda_F$ are the MSSM gauginos
and the flavor $U(1)_A$ gaugino. ($M^i_{1/2}$, $M_{\lambda_F}$)
and ($(\tilde{m}^2_f)_{ab}$, $\tilde{m}_s^2$) are the gaugino and
scalar soft masses respectively. $\tilde{f}_a$ stands for the MSSM
sfermions including the right--handed sneutrinos. Note that the
generalized $A$--terms in Eq. (\ref{lagr1}) has the same structure
as the corresponding superpotential terms of Eq. (\ref{superP1}).

We assign flavor $U(1)_A$ charges to the MSSM fields such that the
observed fermion mass and mixing hierarchies are obtained with all
Yukawa couplings being order one. As we will show explicitly, the
expansion parameter $\e= \langle S\rangle/M_{st}$ is naturally of
order $0.2$ in anomalous $U(1)$ models. We use the idea of
``lopsided" mass matrices for generating large neutrino mixings
\cite{bb, Albright}, while maintaining small quark mixings.

We analyze two specific textures for the fermion mass matrices,
Model 1 and Model 2. Model 2 is the texture advocated in Ref.
\cite{bgw}, and Model 1 is a slight variant of it. The textures
for the two models are
\begin{eqnarray}\label{massM1}
&&M_u\sim \langle
H_u\rangle\pmatrix{\epsilon^{\,8-2\alpha}&\epsilon^{\,6-\alpha}&\epsilon^{\,4-\alpha}\cr
\epsilon^{\,6-\alpha}&\epsilon^4&\epsilon^2\cr\epsilon^{\,4-\alpha}&\epsilon^2&1}\,,\hspace{1.cm}
M_d\sim \langle H_d\rangle\epsilon^p
\pmatrix{\epsilon^{\,5-\alpha}&\epsilon^{\,4-\alpha}&\epsilon^{\,4-\alpha}\cr
\epsilon^3 &\epsilon^2&\epsilon^2\cr\epsilon&1&1},\nn\\
\nn\\
\vspace{.5cm} &&M_e\sim \langle
H_d\rangle\epsilon^p\pmatrix{\epsilon^{\,5-\alpha}&\epsilon^3&\epsilon\cr
\epsilon^{\,4-\alpha}&\epsilon^2&1\cr\epsilon^{\,4-\alpha}&\epsilon^2&1}\,,\hspace{2.cm}
M_{\nu_D}\sim \langle H_u\rangle\epsilon^s
\pmatrix{\epsilon^2&\epsilon &\epsilon \cr \epsilon
&1&1\cr\epsilon&1&1},\nn\\
\nn\\
 \vspace{.5cm} &&M_{\nu^c}\sim M_R
\pmatrix{\epsilon^2&\epsilon&\epsilon\cr
\epsilon&1&1\cr\epsilon&1&1}\,\hspace{1.2cm}\Rightarrow
\hspace{1.5cm} M^{light}_\nu\sim \frac{{\langle
H_u\rangle}^2}{M_R}\epsilon^{2s}
\pmatrix{\epsilon^2&\epsilon&\epsilon\cr
\epsilon&1&1\cr\epsilon&1&1}.
\end{eqnarray}
Here $M_u$, $M_d$ and $M_e$ are the up--quark, down--quark, and
the charged lepton mass matrices (written in the basis $uM_uu^c$,
etc.). $M_{\nu_D}$ is the Dirac neutrino mass matrix, and
$M_{\nu^c}$ is the right--handed neutrino Majorana mass matrix.
The light neutrino mass matrix $M^{light}_\nu$ is derived from the
seesaw mechanism. We have not exhibited order one coefficients in
the matrix elements of Eq. (\ref{massM1}). The expansion parameter
is $\e\sim 0.2$. The exponent $p$ appearing in the overall factor
$\e^p$ multiplying $M_d$ and $M_e$ is assumed to take values $0$,
$1$ or $2$ corresponding to large ($\sim 20$), moderate ($\sim
10$), and small ($\sim 5$) values of $\tan\beta\,\,(\equiv \langle
H_u\rangle/\langle H_d\rangle)$ respectively.

The parameter $\alpha$ is allowed to take two values, $0$ and $1$,
corresponding to Model 1 ($\alpha=0$) and Model 2 ($\alpha=1$).
The two models differ only in the masses and mixings of the first
family. Both models give excellent fits to the fermion masses and
mixings including neutrino oscillation parameters. Their
predictions for LFV are however noticeably different, which we
analyze in Section 4.

\subsection{Anomalous $U(1)_A$ Models}

The mass matrix textures of Eq. (\ref{massM1}) can be obtained
from an anomalous $U(1)$ gauge symmetry that is flavor dependent.
In Table \ref{tcharge} we present a set of such $U(1)$ charges.
These charges are integer multiples of the charge of the flavon
field $S$ which is taken to be $-1$. Note that the texture alone
does not fix the exponent $s$ appearing in $M_{\nu_D}$ of Eq.
(\ref{massM1}).
\begin{table}[h]
\begin{center}
\begin{tabular}{|c|c|c|}\hline
\rule[5mm]{0mm}{0pt}Field& $U(1)_A$ Charge& Charge
notation\\\hline
\rule[5mm]{0mm}{0pt}$Q_1$, $Q_2$, $Q_3$& \,\,$ 4-\alpha, \,2, \,0$ &$q^Q_i$\\
\rule[5mm]{0mm}{0pt}$L_1$, $L_2$, $L_3$&\,\,$ 1+s,\, s,
\,s$&$q^L_i$\\
\rule[5mm]{0mm}{0pt}$u^c_1$, $u^c_2$, $u^c_3$&$ 4-\alpha,\, 2,\, 0$&$q^u_i$\\
\rule[5mm]{0mm}{0pt}$d^c_1$, $d^c_2$, $d^c_3$&$1+p,\, p,\,
p$& $q^d_i$\\
\rule[5mm]{0mm}{0pt}$e^c_1$,$e^c_2$,$e^c_3$&\,\,$ 4+p-s-\alpha,\, 2+p-s,\, p-s$\,\,&$q^e_i$\\
\rule[5mm]{0mm}{0pt}$\nu^c_1$, $\nu^c_2$,
$\nu^c_3$ &$ 1,\, 0,\, 0$&$q^\nu_i$\\
\rule[5mm]{0mm}{0pt}$H_u$, $H_d$, $S$&$ 0,\, 0,\, -1$&$(h, \bar{h}, q_s)$\\
\hline
\end{tabular}
\caption{\footnotesize The flavor $U(1)_A$ charge assignments for
the MSSM fields and the flavon field $S$ in the normalization of
$q_s=-1$. Here $\alpha$ is $0\,(1)$ for Model 1 (Model 2). In the
third column we list the generic notation for the charges used in
the RGE analysis.}
  \label{tcharge}
\end{center}
\end{table}

We use the Green--Schwarz (GS) mechanism \cite{GS} for anomaly
cancellation associated with $U(1)_A$ gauge symmetry. These
conditions are given by
\begin{eqnarray}\label{GS1}
\frac{A_1}{k_1}=\frac{A_2}{k_2}=\frac{A_3}{k_3}=\frac{A_{A}}{k_A}=\frac{A_{gravity}}{12}\,,
\end{eqnarray}
where $A_i$ and $A_A$ are the coefficients  $U(1)_Y^2\times
U(1)_A$, $SU(2)_L^2\times U(1)_A$, $SU(3)_C^2\times U(1)_A$ and
$U(1)_A^3$ gauge anomalies respectively. $k_i$ and $k_A$ are the
Kac--Moody levels. $A_{gravity}$ is the mixed gravitational
anomaly coefficient  which is given by the trace of the $U(1)_A$
charges over all fields. With the non--Abelian levels $k_2=k_3=1$,
which is the simplest possibility, from Table \ref{tcharge} and
Eq. (\ref{GS1}) one finds
\begin{eqnarray}\label{Anom1}
A_2&=&\frac{19-3\alpha+3s}{2},\nn\\
A_3&=&\frac{19-3\alpha+3p}{2}\,.
\end{eqnarray}
This implies that $p=s$. Furthermore,
\begin{eqnarray}\label{Anom2}
A_1&=&\frac{5}{6}(19-3\alpha+3p)\,,
\end{eqnarray}
which fixes the  level $k_1$ to be $5/3$.

With $p=s$  the charges given in Table \ref{tcharge} are
compatible with $SU(5)$ unification. In string theory gauge
coupling unification can occur without a simple covering group.
The string unification condition is \cite{Gins}
\begin{eqnarray}\label{Norml}
k_1g_1^2=k_2g_2^2=k_3g_3^2=k_Ag_F^2\,.
\end{eqnarray}
Our result $k_1=5/3$ is what is needed for consistency of the
observed unification of gauge couplings in the MSSM. The
discrepancy in the unification scale derived from low energy data
versus perturbative string theory evaluation can be reconciled in
the context of M--theory by making use of the radius of the
eleventh dimension \cite{witten}. We assume such a scenario.

From now on we shall assume the $SU(5)$ normalization for $g_1$.
If one assumes that the field content of the model is just the one
listed in Table \ref{tcharge}, the gravitational anomaly
$A_{gravity}$ would not satisfy the GS condition. One simple
solution is the introduction of additional MSSM singlet (hidden
sector) fields $X_k$. Then Eq. (\ref{GS1}) leads to the following
result:
\begin{eqnarray}
A_{gravity}=5\left(13-2\alpha+3p\right)+\sum_kq^X_k=6(19-3\alpha+3p)\,,
\end{eqnarray}
where $q^X_k$ are the $U(1)_A$ charges of the extra fields $X_k$.
From this, one gets $ \sum_kq^X_k=\left(49-8\alpha+3p\right)$. We
assume for simplicity that all the $X_k$ fields have the same
flavor charge equal to $1$.  The number $n^X$ of $X_k$ fields is
then  fixed to be
\begin{eqnarray}\label{nx}
n^X=49-8\alpha+3p\,.
\end{eqnarray}

We are now in a position to determine the level $k_A$ as well as
the $U(1)_A$ gauge coupling $g_F$ at the unification scale. We
renormalize the $U(1)_A$ charges by a factor $|q_s|$ so that the
charge of the flavon field is now $-|q_s|$. $|q_s|$ is determined
by demanding $g_F^2=g_2^2$ at the unification scale. Eq.
(\ref{GS1}) and the number $n^X$ in Eq. (\ref{nx}) then fix
$|q_s|$ to be
\begin{eqnarray}\label{norm}
|q_s|=\sqrt{\frac{19-3\alpha+3p}{10(4-\alpha)^3+5[(1+p)^3+2p^3]+n^X}}\,.
\end{eqnarray}
For  $p=(0,\,1,\,2)$ one has $|q_s|=(0.165,\,0.172,\,0.166)$ for
Model 1 ($\alpha=0$) and $|q_s|=(0.225,\,0.228,\,0.203)$ for Model
2 ($\alpha=1$)\,.

The Fayet--Iliopoulos term for the anomalous $U(1)_A$, generated
through the gravitational anomaly, is given by \cite{DSW}
\begin{eqnarray}
\xi=\frac{g_{st}^2M_{st}^2}{192\pi^2}|q_s|A_{gravity}\,,
\end{eqnarray}
where $g_{st}$ is the unified gauge coupling at the string scale.
By minimizing the potential
\begin{eqnarray}\label{Dterm}
V=\frac{|q_s|^2g_F^2}{8}\left(\frac{\xi}{|q_s|}-|S|^2+\sum_aq_a^f|\tilde{f}_a|^2+\sum_k
q_k^X|X_k|^2\right)^2
\end{eqnarray}
in the unbroken supersymmetric limit we obtain
\begin{eqnarray}\label{13}
\epsilon=\frac{\langle
S\rangle}{M_{st}}=\sqrt{\frac{g_{st}^2A_{gravity}}{192\pi^2}}\,.
\end{eqnarray}
The numerical values of $\e$ derived from Eq. (\ref{13})  for
different $p$ and $\alpha$ are listed in Table \ref{eps1} by
making use of Eq. (\ref{norm}).
\begin{table}
\begin{center}
\begin{tabular}{|c|c|c|c|}\hline
\rule[5mm]{0mm}{0pt}$\e$&$p=0$&$ p=1$&$ p=2$\\\hline
\rule[5mm]{0mm}{0pt}$\alpha=0$&$0.177$& $0.191$& $0.204$\\\hline
\rule[5mm]{0mm}{0pt}$\alpha=1$&$0.163$& $0.177$& $0.191$\\\hline
\end{tabular}
\caption{\footnotesize Numerical values for the small expansion
parameter $\e$ corresponding to different fermion mass hierarchy
structure.}
  \label{eps1}
\end{center}
\end{table}
This is the small expansion parameter appearing in the mass
matrices of Eq. (\ref{massM1}). Here we took $g^2_{st}/4\pi\simeq
1/24$.

The mass of the flavor $U(1)_A$ gauge boson is found to be
\begin{eqnarray}\label{MA1}
M_F=\frac{|q_s|g_F\langle S\rangle}{\sqrt{2}}\,.
\end{eqnarray}
Between the string scale $M_{st}$ and $M_F$ the flavor gaugino
contributes to flavor violating processes. This mass can now be
determined:
\begin{eqnarray}\label{MAM1}
M_F=\left(\frac{M_{st}}{88.3},\,
\frac{M_{st}}{84.6},\,\frac{M_{st}}{86.3}\right)\, \mbox{for
$p=(0, 1, 2$)}\, ,
\end{eqnarray}
in the case of Model 1 and
\begin{eqnarray}\label{MAM2}
M_F=\left(\frac{M_{st}}{68.7},\,
\frac{M_{st}}{66.8},\,\frac{M_{st}}{72.9}\right)\, \mbox{for
$p=(0, 1, 2$)}\, ,
\end{eqnarray}
in the case of Model 2.

\subsection{Fermion Mass Fits}

Here we present numerical fits to the fermion masses and mixings
for Model 1 and Model 2. These fits will be used in our
quantitative analysis of lepton flavor violation.

As input at low energy we choose the following values for the
running quark masses  \cite{gasser}
\begin{eqnarray}
m_u(1\,\mbox{GeV})=5.11\,\mbox{MeV},\,\,\,\,\,\, m_c(m_c)=1.27\,\mbox{GeV},\,\,\,\,\,\, m_t(m_t)=167\,\mbox{GeV},\nn\\
m_d(1\,\mbox{GeV})=8.9\,\mbox{MeV},\,\,\,\,\,\,m_s(1\,\mbox{GeV})=130\,\mbox{MeV},\,\,\,\,\,\,m_b(m_b)=4.25\,\mbox{GeV}.
\end{eqnarray}
The CKM mixing matrix elements are chosen to be $|V_{us}|=0.222$,
$|V_{ub}|=0.0035$, $|V_{cb}|=0.04$ and $\eta=0.33$ (the
Wolfenstein parameter of CP--violation). Using two--loop QED and
QCD renormalization group equations we obtain these running
parameters at the SUSY breaking scale, $M_{SUSY}=500$ GeV, with
$\alpha_s(M_Z)=0.118$, to be
\begin{eqnarray}
r_f\equiv\frac{m_f(M_{SUSY})}{m_f(m_f)}\, ,
\end{eqnarray}
where
\begin{eqnarray}
(r_t,\, r_b,\, r_\tau,\, r_{u,c},\, r_{d,s},\,
r_{e,\mu})=\left(0.943,\, 0.605,\, 0.991,\, 0.395,\, 0.398,\,
0.989\right).
\end{eqnarray}
Using two--loop SUSY RGE evaluation above $M_{SUSY}$ we obtain the
Yukawa couplings at the $U(1)_A$ breaking scale ($\sim10^{15}$
GeV) to be
\begin{eqnarray}\label{Yuk1}
&&\left(Y_u,\,Y_c,\,Y_t\right)=\left(5.135\times10^{-6},\,1.426\times10^{-3},\,0.538\right),\nn\\
&&\left(Y_d,\,Y_s,\,Y_b\right)=\left(3.459\times10^{-5},\,5.052\times10^{-4},\,2.768\times10^{-2}\right),\nn\\
&&\left(Y_e,\,Y_\mu,\,Y_\tau\right)=\left(1.024\times10^{-5},\,2.118\times10^{-3},\,3.572\times10^{-2}\right),\nn\\
&&\left(Y_{\nu_1},\,Y_{\nu_2},\,Y_{\nu_3}\right)=\left(3.515\times10^{-4},\,8.419\times10^{-4},\,1.131\times10^{-2}\right),
\end{eqnarray}
for $\tan\beta=5$,
\begin{eqnarray}\label{Yuk2}
&&\left(Y_u,\,Y_c,\,Y_t\right)=\left(4.999\times10^{-6},\,1.389\times10^{-3},\,0.518\right),\nn\\
&&\left(Y_d,\,Y_s,\,Y_b\right)=\left(6.844\times10^{-5},\,9.997\times10^{-4},\,5.470\times10^{-2}\right),\nn\\
&&\left(Y_e,\,Y_\mu,\,Y_\tau\right)=\left(2.027\times10^{-5},\,4.192\times10^{-3},\,7.094\times10^{-2}\right),\nn\\
&&\left(Y_{\nu_1},\,Y_{\nu_2},\,Y_{\nu_3}\right)=\left(1.708\times10^{-3},\,4.105\times10^{-2},\,5.519\times10^{-2}\right),
\end{eqnarray}
for $\tan\beta=10$, and
\begin{eqnarray}\label{Yuk3}
&&\left(Y_u,\,Y_c,\,Y_t\right)=\left(4.996\times10^{-6},\,1.387\times10^{-3},\,0.518\right),\nn\\
&&\left(Y_d,\,Y_s,\,Y_b\right)=\left(1.40\times10^{-4},\,2.045\times10^{-4},\,0.113\right),\nn\\
&&\left(Y_e,\,Y_\mu,\,Y_\tau\right)=\left(4.132\times10^{-5},\,8.545\times10^{-3},\,0.147\right),\nn\\
&&\left(Y_{\nu_1},\,Y_{\nu_2},\,Y_{\nu_3}\right)=\left(8.551\times10^{-3},\,2.059\times10^{-2},\,0.278\right),
\end{eqnarray}
for $\tan\beta=20$. $|V_{ub}|$, $|V_{cb}|$, $|V_{td}|$ and
$|V_{ts}|$ are multiplicatively renormalized  by an RGE factor of
$0.9$ in going from the low energy scale to the $U(1)_A$ breaking
scale.

We have determined the Dirac neutrino Yukawa couplings as follows.
First we note that the anomaly cancellation conditions in Eq.
(\ref{GS1}) implies $p=s$, which means that the Dirac neutrino
Yukawa couplings are fixed to be of the same order as the charged
lepton Yukawa couplings. Now, if one takes the right--handed
Majorana neutrino mass matrix to be proportional to the transpose
of the Dirac neutrino Yukawa coupling matrix for simplicity,
$M_{\nu^c}=Y_{\nu}^T\,M_R\,\e^p$, then the light neutrino mass
matrix is given by
\begin{eqnarray}
M^{light}_{\nu}=Y_{\nu}M_{\nu^c}^{-1}Y_{\nu}^T\,v^2\,sin^2\beta=Y_{\nu}\frac{v^2\,sin^2\beta}{M_R\,\e^p}\,.
\end{eqnarray}
This simplified choice is certainly consistent with the fermion
mass structures we have chosen in Eqs. (\ref{massM1}). We adopt
this choice in our analysis. $Y^\nu$ is determined from a fit to
the light neutrino oscillation parameters with $M_R=10^{14}$ GeV.
This fit corresponds to $m_{\nu_e}=2.7\times10^{-3} $ eV, $
m_{\nu_{\mu}}=6.4\times10^{-3}$ eV and
$m_{\nu_{\tau}}=8.6\times10^{-2}$ eV\, and the leptonic mixing
matrix given by
\begin{eqnarray}\label{MNS}
V_{MNS}=\pmatrix{0.848&-0.526&-0.0409\cr
    0.349&0.619&-0.72\cr
    -0.4&-0.59&-0.7013}.
\end{eqnarray}
We also consider a scenario where the Dirac neutrino Yukawa
couplings are maximized by choosing $M_R=4\times10^{14}$ GeV. In
this case we have
\begin{eqnarray}\label{Yuk4}
&&\left(Y_{\nu_1},\,Y_{\nu_2},\,Y_{\nu_3}\right)=\left(1.406\times10^{-3},\,3.368\times10^{-3},\,0.453\times10^{-2}\right)\,\mbox{for}\,\tan\beta=5,\nn\\
&&\left(Y_{\nu_1},\,Y_{\nu_2},\,Y_{\nu_3}\right)=\left(6.843\times10^{-3},\,1.645\times10^{-2},\,0.222\right)\,\mbox{for}\,\tan\beta=10,\nn\\
&&\left(Y_{\nu_1},\,Y_{\nu_2},\,Y_{\nu_3}\right)=\left(3.514\times10^{-2},\,8.464\times10^{-2},\,1.237\right)\,\mbox{for}\,\tan\beta=20.
\end{eqnarray}

We now present our fits to the observables of Eqs.
(\ref{Yuk1})--(\ref{Yuk4}) consistent with the texture of Eq.
(\ref{massM1}). This cannot be done uniquely since the
right--handed rotation matrices are unknown from low energy data,
so we make a specific choice. In our lepton flavor violation
analysis we shall make use of this specific fit. One should bear
in mind that there are uncertain coefficients of order one in the
Yukawa matrices of our fit, which can lead to an order of
magnitude uncertainty in the branching ratios for LFV processes.

We introduce the following notation:
\begin{eqnarray}
&&Y^f_{ij}\equiv y^f_{ij}\,\e^{n^f_{ij}}.
\end{eqnarray}
In Model 1, a good fit to the Yukawa couplings matrices is found
to be
\begin{eqnarray*}
Y^u=y^u_{33}\pmatrix{3.91\,\epsilon^8&0.226\,\epsilon^6&0.375\,\epsilon^4\cr
0.226\,\epsilon^6&1.91\,\epsilon^4&0.499\,\epsilon^2\cr0.375\,\epsilon^4&0.499\,\epsilon^2&1},\hspace{.5cm}
\end{eqnarray*}
\begin{eqnarray*}
Y^d=y^d_{33}\,\epsilon^p
\pmatrix{(1.56+0.115i)\,\epsilon^5&(0.909+0.054i)\epsilon^4&(0.658+0.131i)\,\epsilon^4\cr
-2.89\,\epsilon^3
&1.02\,\epsilon^2&1.22\,\epsilon^2\cr(-0.878+0.88\times
10^{-3}i)\,\epsilon&0.412+0.11\times10^{-6}i&1+0.73\times10^{-6}i},
\end{eqnarray*}
\begin{eqnarray*}
 Y^e=
y^e_{33}\,\epsilon^p\pmatrix{1.89\,\epsilon^5&1.57\,\epsilon^3&0.812\,\epsilon\cr
0.487\,\epsilon^4&2.14\,\epsilon^2&0.316\cr1.10\,\epsilon^4&1.52\,\epsilon^2&1},\hspace{.5cm}
\end{eqnarray*}
\begin{eqnarray}\label{yukmod11}
Y^{\nu}=y^{\nu}_{33}\, \epsilon^p
\pmatrix{1.51\,\epsilon^2&-0.358\,\epsilon &-0.438\,\epsilon \cr
-0.358\,\epsilon &0.339&0.485\cr-0.438\,\epsilon&0.485&1}.
\end{eqnarray}
Here
\begin{eqnarray}\label{yukmod12}
&&y^u_{33} = \left(\,0.539,\,0.523,\,0.519\right),\nn\\
&&y^d_{33} = \left(\,0.650, \,0.257,\, 0.106\right),\nn\\
&&y^e_{33} = \left(\,0.840, \,0.333,\, 0.139\right),\nn\\
&&y^{\nu}_{33} = \left(\,0.225, \,0.219, \,0.221\right),
\end{eqnarray}
for ($p=2,\, 1,\, 0$) which we shall associate with
$\tan\beta=(\,5,\,10,\,20)$. Here we have taken $\e=0.2$. For
simplicity we assumed the leptonic Yukawa couplings to be all
real.

In Model 2 we have the following fit for the  Yukawa coupling
matrices:
\begin{eqnarray*}
Y^u=
y^u_{33}\pmatrix{0.876\,\epsilon^6&1.30\,\epsilon^5&0.499\,\epsilon^3\cr
1.30\,\epsilon^5&2.59\,\epsilon^4&0.993\,\epsilon^2\cr0.499\,\epsilon^3&0.993\,\epsilon^2&1},\hspace{.5cm}
\end{eqnarray*}
\begin{eqnarray*}
Y^d=y^d_{33}\,\epsilon^p
\pmatrix{(3.01+0.13i)\,\epsilon^4&(2.66+0.13i)\epsilon^3&(1.21+0.13i)\,\epsilon^3\cr
1.79\,\epsilon^3
&2.26\,\epsilon^2&1.42\,\epsilon^2\cr(1.00+0.33\times
10^{-3}i)\,\epsilon&0.987+0.582\times10^{-5}i&1+0.21\times10^{-5}i},
\end{eqnarray*}
\begin{eqnarray*}
 Y^e=
y^e_{33}\,\epsilon^p\pmatrix{1.19\,\epsilon^4&1.68\,\epsilon^3&0.579\,\epsilon\cr
0.892\,\epsilon^3&2.18\,\epsilon^2&0.350\cr1.36\,\epsilon^3&1.45\,\epsilon^2&1},\hspace{.5cm}
\end{eqnarray*}
\begin{eqnarray}\label{yukmod21}
Y^{\nu}=y^{\nu}_{33}\, \epsilon^p
\pmatrix{1.53\,\epsilon^2&-0.329\,\epsilon &-0.406\,\epsilon \cr
-0.329\,\epsilon &0.293&0.449\cr-0.406\,\epsilon&0.449&1}.
\end{eqnarray}
Here
\begin{eqnarray}\label{yukmod22}
&&y^u_{33} = \left(\,0.535,\,0.52,\,0.515\right),\nn\\
&&y^d_{33} = \left(\,0.650,\, 0.257,\, 0.107\right),\nn\\
&&y^e_{33} = \left(\,0.840, \,0.333, \,0.139\right),\nn\\
&&y^{\nu}_{33} = \left(\,0.233, \,0.228,\, 0.229\right),
\end{eqnarray}
for three different values of $p=(2,\, 1,\, 0)$ identified with
$\tan\beta=(\,5,\,10,\,20)$.

\subsection{Leptogenesis Constraints}

In this subsection we show that baryogenesis through leptogenesis
\cite{FY}  can occur naturally in our models of fermion mass
hierarchy. This requirement does put  significant restrictions on
the parameter $p$, which is related to the value of $\tan\beta$.
We find that an acceptable baryon asymmetry would require $p = 2,
1$, corresponding to small to moderate values of $\tan\beta$.  In
this case, lepton flavor violation resulting from the
right--handed neutrino Dirac Yukawa couplings tend to be
suppressed.

From the neutrino Majorana mass matrix of Eq. (3), we find that
the heavy Majorana masses of $\nu_{2,3}^c$ denoted by $M_{2,3}$
are of order $M_R$, while that of $\nu_1^c$ (denoted by $M_1$) is
of order $\epsilon^2\, M_R$. Lepton asymmetry is generated in the
out of equilibrium decay of $\nu^c_1$.  The induced asymmetry
parameter is given by
\begin{eqnarray}
\e_1=\frac{1}{8\pi v_u^2}\frac{1}{\left(M_{\nu_D}^\dagger
M_{\nu_D}\right)_{11}}\sum_{j=2,3}
\mbox{Im}\left[\left(M_{\nu_D}^\dagger
M_{\nu_D}\right)^2_{1j}\right]f\left(\frac{M^2_j}{M^2_1}\right)\,,
\end{eqnarray}
where $v_u \simeq 174\sin\beta$ GeV, and the function $f(x) \simeq
-3/\sqrt{x}$.  Let us focus on the asymmetry induced through the
exchange of $\nu_2^c$. Denote the mass ratio $M_1/M_2 = a
\epsilon^2$, where $a$ is a coefficient of order one. We also have
from Eq. (3) $(M_{\nu_D}^\dagger M_{\nu_D})_{11} = b \e^{2s+2}
v_u^2\,$, $(M_{\nu_D}^\dagger M_{\nu_D})_{12} = c \e^{2s+1}
v_u^2\,$, where $b,\,c$ are also order one coefficients.
$\epsilon_1$ can then be estimated to be
\begin{eqnarray}
\e_1\simeq \frac{3ac}{8\pi b}\e^{2s+2}\sin\phi\,,
\end{eqnarray}
where $\sin\phi$ is an order one phase parameter. The leptonic
asymmetry parameter $Y_L$ is obtained from $\epsilon_1$ via the
relation $Y_L = d \epsilon_1/g^*$, where $d$ is the dilution
factor and $g^*$ is the number of degrees of freedom in thermal
equilibrium ($g^* \simeq 200$ in the supersymmetric Standard
Model).  The dilution factor $d$ depends on the ratio $k$ of the
decay rate versus the Hubble expansion rate \cite{param}:
\begin{eqnarray}
k = \frac{M_{Pl}\left(M_{\nu_D}^\dagger
M_{\nu_D}\right)_{11}}{1.66\sqrt{g^*}(8\pi v^2_u)M_1}\,.
\end{eqnarray}

We now demand that the light neutrino mass $m_{\nu_3}$, which can
be written from the last expression of Eq. (\ref{massM1}) as
$m_{\nu_3} = h \e^{2s} v_u^2/M_2\,$, should equal $0.05$ eV, to be
consistent with the atmospheric neutrino oscillation data.  This
then determines $k$ to be $k \simeq [b/(ah)](33.3)\,$. For this
range of $k$ (assuming that $b/(ah)$ is not smaller than about
$0.3$) the dilution factor can be written approximately as
\begin{eqnarray}
d \simeq {0.3 \over k [{\rm ln}(k)]^{0.6}},~~( 10 \leq k \leq
10^6).
\end{eqnarray}
Using $Y_B \simeq -Y_L/2$, we find
\begin{eqnarray}
Y_B = {3.6 \times 10^{-6} x \e^{2s} \over [{\rm ln}(33.3
 y)]^{0.6}}\,,
\end{eqnarray}
where $x = (a^2 c^2 h\sin\phi)/b^2$ and $y = b/(ah)$ are both
expected to be of order one.  For the case of $s = 2\,$, we choose
$x=0.05,\, y=3$ to obtain $Y_B = 1 \times 10^{-10}$, which is near
the central value of the observed baryon asymmetry.  Note that
this choice of $x,\,y$ is quite natural and  consistent with the
flavor structure we have adopted. For the case of $s=1\,$, we can
fit $Y_B = 1 \times 10^{-10}$ by choosing $x=1.7\times 10^{-3},\,
y=3\,$. Considering that $x$ is a nontrivial combination of order
one parameters, this choice of $x$ cannot be considered unnatural.
On the other hand, for the case of $s=0\,$, a good fit to $Y_B$
requires, for example, $x= 7 \times 10^{-5},\, y=3$. Such a low
value of $x$ does not go well with the hierarchical structure
dictated by the anomalous $U(1)_A$ symmetry.

We conclude that baryogenesis via leptogenesis works in a simple
and natural way with the mass matrix textures suggested in Eq.
(3), but only for low to medium values of $\tan\beta\,$,
corresponding to $s=p = 1,\, 2\,$.

\section{Generalized RGE Analysis of Soft SUSY Breaking Parameters}

In this section we give a general RGE analysis of the soft SUSY
breaking parameters including higher dimensional operators as
shown in Eqs. (\ref{superP1}) and (\ref{lagr1}). This includes the
effects of the flavor $U(1)_A$ gaugino sector. Our analysis of
this section should apply to a large class of Froggatt-Nielsen
models.

The one--loop $\beta$--functions for the soft scalar masses of the
sleptons are found to be
\begin{eqnarray}\label{betams2}
\beta\left({\tilde{m}_L}^2\right)_{ij}&=&\beta\left({\tilde{m}_L}^2\right)^{MSSM}_{ij}+\frac{1}{16\pi^2}\left\{\left(\tilde{m}_L^2{Y^\nu}^\dagger
Y^\nu +{Y^\nu}^\dagger
Y^\nu{\tilde{m}_{L}}^2\right)_{ij}\right.\nn\\&&\hspace{1cm}+2\left({Y^\nu}^\dagger\tilde{m}_\nu^2Y^\nu+\tilde{m}_{H_u}^2{Y^\nu}^\dagger
Y^\nu+{A^\nu}^\dagger
A^\nu\right)_{ij}\nn\\&&\left.\hspace{1cm}+2q^L_ig_F^2\delta_{ij}\left(\sigma-4q^L_i(M_{\lambda_F})^2\right)\right\}\,,
\end{eqnarray}
\begin{eqnarray}\label{betams3}
\beta\left({\tilde{m}_e}^2\right)_{ij}&=&\beta\left({\tilde{m}_e}^2\right)^{MSSM}_{ij}+\frac{1}{16\pi^2}
2q^e_ig_F^2\delta_{ij}\left(\sigma-4q^e_i(M_{\lambda_F})^2\right),
\end{eqnarray}
\begin{eqnarray}\label{betams4}
\beta\left({\tilde{m}_\nu}^2\right)_{ij}&=&\frac{1}{16\pi^2}\left\{2\left(\tilde{m}_\nu^2Y^\nu
{Y^\nu}^\dagger +Y^\nu
{Y^\nu}^\dagger{\tilde{m}_{\nu}}^2\right)_{ij}\right.\nn\\&&\hspace{1cm}+4\left({Y^\nu}\tilde{m}_\nu^2{Y^\nu}^\dagger+\tilde{m}_{H_u}^2Y^\nu
{Y^\nu}^\dagger+A^\nu {A^\nu}^\dagger\right)_{ij}\nn\\&&\left.
\hspace{1cm}+2q^\nu_ig_F^2\delta_{ij}\left(\sigma-4q^\nu_i(M_{\lambda_F})^2\right)\right\}.
\end{eqnarray}
Similarly the $\beta$--functions for the squark soft masses are
given by
\begin{eqnarray}\label{betams1}
\beta\left({\tilde{m}_Q}^2\right)_{ij}&=&\beta\left({\tilde{m}_Q}^2\right)^{MSSM}_{ij}+\frac{1}{16\pi^2}
2q^Q_ig_F^2\delta_{ij}\left(\sigma-4q^Q_i(M_{\lambda_F})^2\right),\\
\beta\left({\tilde{m}_u}^2\right)_{ij}&=&\beta\left({\tilde{m}_u}^2\right)^{MSSM}_{ij}+\frac{1}{16\pi^2}
2q^u_ig_F^2\delta_{ij}\left(\sigma-4q^u_i(M_{\lambda_F})^2\right),
\end{eqnarray}
\begin{eqnarray}\label{betams5}
\beta\left({\tilde{m}_d}^2\right)_{ij}&=&\beta\left({\tilde{m}_d}^2\right)^{MSSM}_{ij}+\frac{1}{16\pi^2}
2q^d_ig_F^2\delta_{ij}\left(\sigma-4q^d_i(M_{\lambda_F})^2\right)\,.
\end{eqnarray}
Here $\sigma$ is defined as
\begin{eqnarray}\label{sigma}
&&\sigma=3\,{\rm
Tr}\left(2q^Q{\tilde{m}_Q}^2+q^u{\tilde{m}_u}^2+q^d{\tilde{m}_d}^2\right)
+{\rm
Tr}\left(2q^L{\tilde{m}_L}^2+q^e{\tilde{m}_{e}}^2+q^\nu{\tilde{m}_\nu}^2\right)\nn\\&&\hspace{2cm}+q_s\tilde{m}_s^2+\sum_k
q^X_k{\tilde{m}_{X_k }}^2\, ,
\end{eqnarray}
where $\tilde{m}_{X_k}$ is the soft mass of the extra particles
$X_k$ and the trace is over family space. Here
$\beta\left(\tilde{m}^2_L\right)^{MSSM}_{ij}$ stands for the MSSM
$\beta$--function without the $\nu^c$ or the flavor $U(1)_A$
contributions \cite{martinvaughn}.

\begin{figure}[ht]
\par
\begin{center}
\epsfig{file=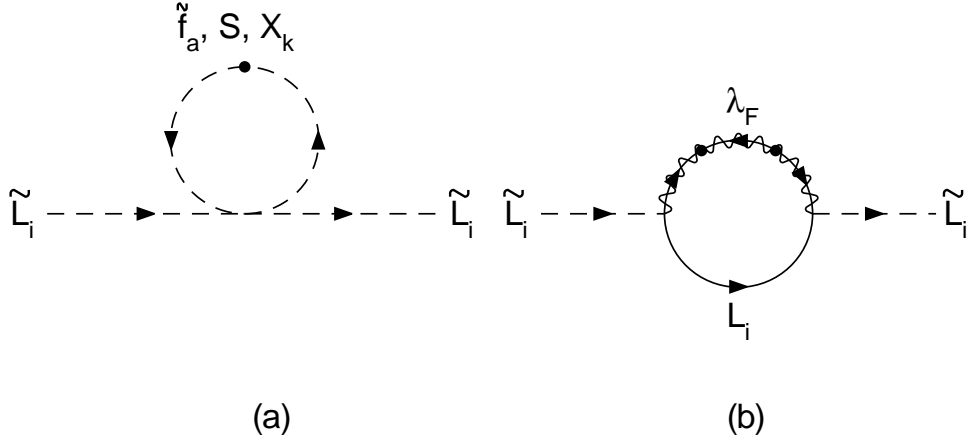,height=2.3in,width=5.in}
\end{center}
 \caption{\footnotesize (a) Trace correction from
$D_A$--term to the soft scalar masses. (b) $U(1)_A$
gaugino--induced corrections to the soft masses.}\label{DiagTrG}
\end{figure}

The contributions proportional to $\sigma$ in Eqs.
(\ref{betams2})--(\ref{betams5}) arise from the diagram in Figure
1 (a) which has its origin from the $U(1)_A$ $D$--term. We call
this the trace contributions. For a non--anomalous $U(1)$ gauge
symmetry with universal scalar masses the trace term would vanish.
However, for an anomalous $U(1)$ gauge symmetry,  trace of the
flavor charges is not zero, so this term will induce flavor
non--universal masses. The diagram in Figure 1 (b) is  the source
of  flavor non--universal contributions proportional to the
gaugino mass $M_{\lambda_F}$ in Eqs.
(\ref{betams2})--(\ref{betams5}).

\begin{figure}[ht]
\par
\begin{center}
\epsfig{file=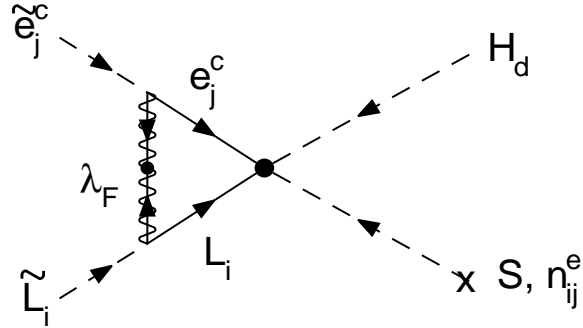,height=1.7in,width=3.in}
\end{center}
\caption{\footnotesize $U(1)_A$ gaugino--induced vertex correction
diagram for $A$--terms.}\label{diagV}
\end{figure}

Now we give the expressions for the one--loop contributions to the
$\beta$--function of the SUSY breaking $A$--terms of Eq.
(\ref{lagr1}). Let us introduce the following notation:
\begin{eqnarray}
A^f_{ij}\equiv a^f_{ij}\,\e^{n^f_{ij}}.
\end{eqnarray}
There are two types of contributions to the $\beta$--functions of
$A^f_{ij}$: one from the gaugino and the other from the
$A$--terms. The flavor gaugino contribution arises from diagrams
such as  the one in  Figure \ref{diagV}. The $A$--term
contribution to $\beta\left(A^f\right)$ cannot have the flavon
field $S$ propagating in the loop, so that contribution is
included in the MSSM piece.

The gaugino vertex contribution to $\beta\left(A^e_{ij}\right)$
(see Figure \ref{diagV}) is
\begin{eqnarray}\label{beta14}
\beta(a^e_{ij})^V=\frac{1}{4\pi^2}M_{\lambda_F}g_F^2y^e_{ij}\left(q^L_i
q^e_j+q^L_i \bar{h} +q^e_j \bar{h} +n^e_{ij}q_s(q^L_i+q^e_j+
\bar{h})+\frac{1}{2}n^e_{ij}(n^e_{ij}-1)q_s^2 \right).
\end{eqnarray}
 Eq. (\ref{beta14}) is
obtained by summing all possible  gaugino exchange diagrams.

We now  list the  full one--loop $\beta$ function for each
$A^f_{ij}$.  This generalizes the results of Martin
\cite{martin1}.
\begin{eqnarray}\label{beta3}
\beta(A^e)_{ij}&=\beta(A^e)^{MSSM}_{ij}+&\frac{1}{16\pi^2}\left\{A^e\left[Y^{\nu^\dagger}Y^\nu-2\left((q^L_i)^2+(q^e_j)^2+\bar{h}^2\right)g_F^2\right]\right.\nn\\
&&\left.\,\,\,\,\,\,\,\,\,\,\,+2Y^eY^{\nu^\dagger}A^\nu\right\}_{ij}+\frac{1}{4\pi^2}g_F^2\,Z^e_{ij}Y^e_{ij}M_{\lambda_F},
\end{eqnarray}
\begin{eqnarray}\label{beta4}
\beta(A^\nu)_{ij}&=&\frac{1}{16\pi^2}\left\{A^\nu\left[5Y^{\nu^\dagger}Y^\nu+Y^{e^\dagger}Y^e+{\rm
Tr}\left(3Y^uY^{u^\dagger}+
Y^\nu Y^{\nu^\dagger}\right)\right.\right.\nn\\
&&\left.\hspace{2cm} -3g_2^2-3g_1^2/5-2\left((q^L_i)^2+(q^\nu_j)^2+h^2\right)g_F^2\,\right]\nn\\
&&\hspace{2cm}+2Y^\nu\left[2Y^{\nu^\dagger}A^\nu+Y^{e^\dagger}A^e
+{\rm Tr}\left(3A^uY^{u^\dagger}+A^\nu Y^{\nu^\dagger}\right)\right.\nn\\
&&\hspace{2cm}\left.\left.+3M_1g_1^2/5+3M_2g_2^2\right]\right\}_{ij}+\frac{1}{4\pi^2}g_F^2\,Z^\nu_{ij}Y^\nu_{ij}M_{\lambda_F}\,,
\end{eqnarray}
\begin{eqnarray}\label{beta1}
\beta(A^u)_{ij}&=&\beta(A^u)^{MSSM}_{ij}+\frac{1}{16\pi^2}\left\{A^u\left[{\rm Tr}\left(Y^\nu Y^{\nu^\dagger}\right)-2\left((q^Q_i)^2+(q^u_j)^2+h^2\right)g_F^2\right]\right.\nn\\
&&\left.\,\,\,\,\,\,\,\,\,\,\,+2Y^u{\rm Tr}\left(A^\nu
Y^{\nu^\dagger}\right)\right\}_{ij}+\frac{1}{4\pi^2}g_F^2\,Z^u_{ij}Y^u_{ij}M_{\lambda_F}\,,
\end{eqnarray}
\begin{eqnarray}\label{beta2}
\beta(A^d)_{ij}&=&\beta(A^d)^{MSSM}_{ij}-\frac{1}{8\pi^2}g_F^2A^d_{ij}\left((q^Q_i)^2+(q^d_j)^2+\bar{h}^2\right)\nn\\
&&\,\,\,\,\,\,\,\,\,\,\,+\frac{1}{4\pi^2}g_F^2\,Z^d_{ij}Y^d_{ij}M_{\lambda_F}\,.
\end{eqnarray}
Here we defined the combination of the $U(1)_A$ charges $Z^f_{ij}$
as
\begin{eqnarray}\label{chargec1}
Z^u_{ij}&=&q^Q_i q^u_j+q^Q_i h +q^u_j h +n^u_{ij}q_s(q^Q_i+q^u_j+
h)+\frac{1}{2}n^u_{ij}(n^u_{ij}-1)q_s^2,\nn\\
Z^d_{ij}&=&q^Q_i q^d_j+q^Q_i \bar{h} +q^d_j \bar{h}
+n^d_{ij}q_s(q^Q_i+q^d_j+
\bar{h})+\frac{1}{2}n^d_{ij}(n^d_{ij}-1)q_s^2,\nn\\
Z^e_{ij}&=&q^L_i q^e_j+q^L_i \bar{h} +q^e_j \bar{h}
+n^e_{ij}q_s(q^L_i+q^e_j+
\bar{h})+\frac{1}{2}n^e_{ij}(n^e_{ij}-1)q_s^2,\nn\\
Z^\nu_{ij}&=&q^L_i q^\nu_j+q^L_i h +q^\nu_j h
+n^{\nu}_{ij}q_s(q^L_i+q^\nu_j+
h)+\frac{1}{2}n^\nu_{ij}(n^\nu_{ij}-1)q_s^2.
\end{eqnarray}
From the charges listed in Table \ref{tcharge} we have
\begin{eqnarray}\label{Ze1}
&&Z^e=-\pmatrix{(\,11,\,13,\,16)\,&(\,4,\,6,\,9)&(\,1,\,3,\,6)\cr
(\,10,\,11,\,13)&(\,3,\,4,\,6)&(\,0,\,1,\,3)\cr
(\,10,\,11,\,13)&(\,3,\,4,\,6)&(\,0,\,1,\,3)}\
\end{eqnarray}
in Model 1 ($\alpha=0$) and
\begin{eqnarray}\label{Ze2}
&&Z^e=-\pmatrix{(\,7,\,9,\,12)\,&(\,4,\,6,\,9)&(\,1,\,3,\,6)\cr
(\,6,\,7,\,9)&(\,3,\,4,\,6)&(\,0,\,1,\,3)\cr
(\,6,\,7,\,9)&(\,3,\,4,\,6)&(\,0,\,1,\,3)}\
\end{eqnarray}
in Model 2 ($\alpha=1$) for the three different values of
$p=(0,\,1,\,2)$.

\section{Lepton Flavor Violating Decays $\mu\rightarrow e\gamma$ and $\tau\rightarrow \mu\gamma$}

\subsection{Qualitative Analysis}

In the Standard Model with small neutrino mass, lepton flavor
violating processes are highly suppressed. On the other hand, in
the presence of low energy supersymmetry, LFV effects can be quite
significant. In particular, LFV induced by the right--handed
neutrino Yukawa couplings in the MSSM can lead to $\mu\rightarrow
e\gamma$ and $\tau\rightarrow \mu\gamma$ decay rates near the
current experimental limits \cite{Hisano}--\cite{babudutta}.

Here we focus on flavor violation in leptonic processes. The
slepton soft masses are more sensitive to the $U(1)_A$ gaugino
corrections compared to those in the  squark  sector.  This is
because flavor violation in the squark sector is diluted by the
gluino focusing effects. This is especially so when one considers
the cosmological constraints on the lightest SUSY particle (LSP)
mass. Demanding that the neutralino LSP constitutes an acceptable
cold dark matter imposes the condition $m_0\simeq M_{1/2}/4.4$ in
the context of supergravity models. This condition results from
the coannihillation mechanism \cite{DarkM,babudutta} for diluting
the dark matter density which requires $\tilde{\tau}_R$ mass to be
about $5-15$ GeV above the LSP mass.

The approximate formulae for the sfermion soft masses in terms of
the universal soft scalar mass $m_0$ and the common gaugino mass
$M_{1/2}$ (for small to medium $\tan\beta$) are
\cite{carenawagner1}
\begin{eqnarray}\label{SoftMapprx1}
&&\tilde{m}^2_L\simeq m^2_0+0.52M^2_{1/2},\nn\\
&&\tilde{m}^2_e\simeq m^2_0+0.15M^2_{1/2},\nn\\
&&\tilde{m}^2_Q\simeq m^2_0+6.5M^2_{1/2},\nn\\
&&\tilde{m}^2_u\simeq \tilde{m}^2_d\simeq m^2_0+6.1M^2_{1/2}.
\end{eqnarray}
From these expressions with $m_0\simeq M_{1/2}/4.4$ we see that
the gaugino focusing effects make the squark soft masses much less
sensitive to any flavor violating contributions.

The right--handed neutrino induced LFV effects in our models
depend on the overall factor $\e^p$ in Eq. (\ref{massM1}). These
processes will be suppressed for $p=1,\,2$ corresponding to low
values of $\tan\beta$.  As we noted in Sec. 2.3, $p=2$ is
preferred for leptogenesis, in which case the $\nu^c$ effects are
small.

There are three different sources of LFV in our models: (i) RGE
effects between $M_{st}$ and the $U(1)_A$ symmetry breaking scale
$M_F$ induced by the $U(1)$ gaugino, (ii) RGE effects between
$M_{st}$ and the right--handed neutrino mass scale $M_R$ induced
by the neutrino Dirac Yukawa couplings, and (iii) the $U(1)_A$
$D$--term. Here we discuss only the RGE effects (i) and (ii). We
call them the flavor gaugino induced LFV and $\nu^c$--induced LFV.
(The $D$--term contributions are not included in our numerical
analysis, but are discussed in subsection 4.3.) We give
approximate formulas for these LFV processes by integrating the
relevant $\beta$--functions derived in Section 3.

We adopt the minimal supergravity scenario (mSUGRA) for
supersymmetry breaking. We assume universality of scalar masses
and proportionality of the $A$--terms and the respective Yukawa
couplings at the string scale. Gaugino mass unification is also
assumed.

The various flavor violating effects are summarized below:

(1) Right-handed neutrino contributions to the scalar soft masses
arising from Eq. (\ref{betams2}) proportional to the Dirac
neutrino Yukawa couplings:
\begin{eqnarray}\label{deltams1n}
\delta\left({\tilde{m}_L}^2\right)^{\nu^c}_{ij}\simeq-\left({Y^\nu}^\dagger
Y^\nu\right)_{ij}\left(3\,{m_0}^2+A_0^2\right)\frac{{\rm
ln}\left(M_{st}/M_{R}\right)}{8\pi^2}\,.
\end{eqnarray}

(2) Trace correction from $D_A$--term in Eqs. (\ref{betams2}) and
(\ref{betams3}) from Figure 1(a):
\begin{eqnarray}\label{deltams1a}
&&\delta\left({\tilde{m}_L}^2\right)^{A}_{ij}\simeq
-q^L_i|q_s|g_F^2\delta_{ij}
\left(3\,m_0^2\,\sum_{i=1,3}\left(n^u_{ii}+n^d_{ii}\right)\nn\right.\\&&\,\,\,\,\,\,\,
\left.+m_0^2\,\sum_{i=1,3}\left(n^e_{ii}+n^\nu_{ii}\right) +n^X
m_0^2-\tilde{m}_s^2\right)\frac{{\rm
ln}\left(M_{st}/M_F\right)}{8\pi^2},\nn
\\&&\delta\left({\tilde{m}_e}^2\right)^{A}_{ij}\simeq
-q^e_i|q_s|g_F^2\delta_{ij}
\left(3\,m_0^2\,\sum_{i=1,3}\left(n^u_{ii}+n^d_{ii}\right)\nn\right.\\&&\,\,\,\,\,\,\,
\left.+m_0^2\,\sum_{i=1,3}\left(n^e_{ii}+n^\nu_{ii}\right) +n^X
m_{0}^2-\tilde{m}_s^2\right)\frac{{\rm
ln}\left(M_{st}/M_F\right)}{8\pi^2}\,.
\end{eqnarray}

(3) Gaugino mass correction from Figure 1(b):
\begin{eqnarray}\label{deltams1g}
\delta\left({\tilde{m}_L}^2\right)^{G}_{ij}\simeq
\left(q^L_i\,g_F\right)^2\delta_{ij}\left(M_{\lambda_F}\right)^2\frac{{\rm
ln}\left(M_{st}/M_F\right)}{2\pi^2}\,,\nn\\
\delta\left({\tilde{m}_e}^2\right)^{G}_{ij}\simeq
\left(q^e_i\,g_F\right)^2\delta_{ij}\left(M_{\lambda_F}\right)^2\frac{{\rm
ln}\left(M_{st}/M_F\right)}{2\pi^2}\,.
\end{eqnarray}

(4) Right--handed neutrino induced vertex correction to the
$A^e$--terms (see Eq. (\ref{beta3})):
\begin{eqnarray}\label{da1} \delta A^e_{ij}\simeq-
3A_0\left(Y^e\,{Y^\nu}^\dagger Y^\nu\right)_{ij}\frac{{\rm
ln}\left(M_{st}/M_R\right)}{16\pi^2}\,.
 \end{eqnarray}

(5) Flavor gaugino vertex correction to the $A^e$--terms arising
from Figure 2 (see the last term of Eq. (\ref{beta3})):
\begin{eqnarray}\label{da2} \delta A^e_{ij}\simeq-M_{\lambda_F}{g_F}^2
Y^e_{ij} Z^e_{ij}\frac{{\rm ln}\left(M_{st}/M_F\right)}{4\pi^2}\,.
 \end{eqnarray}

In addition, we have flavor charge dependent wave function
renormalization of the $A$--terms as given in Eq. (\ref{beta3}).
These are however not significant since they are diagonalized
simultaneously with the corresponding Yukawa couplings. On the
other hand, the vertex corrections to the $A$--terms given in Eqs.
(\ref{da1}) and (\ref{da2}) will induce nonproportionality in
going from $M_{st}$ to $M_{F}$.

The matrix elements $Z^e_{ij}$ in Eq. (\ref{da2}) are given in
Eqs. (\ref{Ze1}) and (\ref{Ze2}) for different values of $p$. The
elements in the $(1,2)$ block of $Z^e$ are rather different from
each other, suggesting that the gaugino vertex contributions can
be very important for the process $\mu\rightarrow e \gamma$. On
the other hand, the elements in the second and the third rows are
identical, hence, $A^e_{23}$ and $A^e_{33}$ run at the same rate
as their corresponding Yukawa couplings do in the short momentum
interval. Therefore, this vertex correction for the process
$\tau\rightarrow \mu \gamma$ is always suppressed in models with
the texture of Eq. (\ref{massM1}).

For $\mu\rightarrow e\gamma$ we find that the most dominant effect
is from the flavor gaugino contributions to the soft masses. This
is due to the following reason. It is proportional to the flavor
charge squared and to the flavor gaugino mass squared (recall that
we have $m_0\simeq M_{1/2}/4.4$), both of which are large. On the
other hand, the trace contributions to the soft masses depend
linearly on the flavor charges and are proportional to $m_0^2$,
which make them relatively small although the trace of the
$U(1)_A$ charges itself is large. The right--handed neutrino
contributions are significant only for $p=0$. For other values of
$p$ the $\nu^c$--contributions to the branching ratio for
$l_i\rightarrow l_j\gamma$ is suppressed by $\e^{4p}$.

We find that the gaugino contribution to the
$\tau\rightarrow\mu\gamma$ decay rate is always suppressed since
$\tau_L$ and $\mu_L$ have the same flavor charges and since the
$\tau_R$--$\mu_R$ mixing angle is of order $\e^2$. The only
significant effect to this process is from the right--handed
neutrino effects when $p=0$.

\subsection{Numerical Results}

In this section we present our numerical results for the LFV
processes $\mu\rightarrow e\gamma$ and $\tau\rightarrow
\mu\gamma$. We adopt the mSUGRA scenario for the SUSY breaking
parameters. At the string scale, taken to be $M_{st}=10^{17}$ GeV,
we assume a universal scalar mass $m_0$ and a common gaugino soft
masses $M_{1/2}$. The unified gauge coupling at $2\times10^{16}$
GeV is taken to be $\alpha_G\simeq 1/24$. We assume the $U(1)_A$
gauge coupling $g_F$ to be equal to $g_2$ at the string scale. We
evolve the soft SUSY breaking parameters from $M_{st}$ to the
$U(1)_A$ gaugino mass $M_F\simeq M_{st}/80$ (see Eqs. (\ref{MAM1})
and (\ref{MAM2})). We use the numerical values of the Yukawa
couplings given in Eqs. (\ref{yukmod11})-(\ref{yukmod22}) for this
evolution.

As explained previously, we take $m_0=M_{1/2}/4.4$ so that the
relic abundance of neutralino dark matter can be reproduced
correctly. With this choice we always find the neutralino to be
the LSP with the $\tilde{\tau}_R$ mass higher than the LSP mass by
$5-15$ GeV. We impose radiative electroweak symmetry breaking
condition. The SUSY higgs mass parameter $\mu$ is chosen to be
positive which is favored by $b\rightarrow s\gamma$.

We take $M_{1/2}$ to vary in the range  $250$ GeV to $1$ TeV. The
lower value satisfies the lightest higgs boson mass limit. We
present the results for three different values of
$\tan\beta=(5,\,10,\,20)$. The corresponding values of the
exponent $p$ are taken to be $p=(2,\,1,\,0)$. The results are
presented for two different values of $A_0=(0,\, 300)$ GeV. When
$\tan\beta=20$, the lower limit on $M_{1/2}$ is around $300$ GeV,
or else the radiative electroweak symmetry breaking would fail.
The branching ratios are plotted against universal gaugino mass
$M_{1/2}$ in Figures 3--14.

\begin{figure}[!h]
\vspace*{4.5truecm}
\begin{center}
\epsfig{file=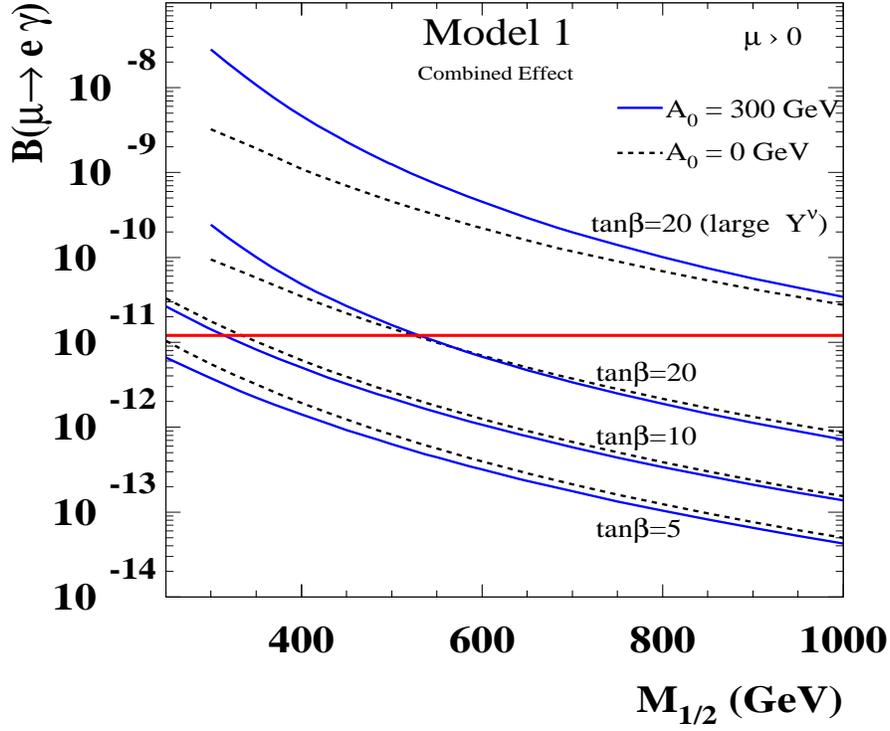,height=3.8in,width=4.6in}
\end{center}
\caption{\footnotesize Branching ratio for the process
$\mu\rightarrow e\gamma$ including all corrections for Model 1.
The solid line corresponds to $A_0=300$ GeV and the dashed line
corresponds to $A_0=0$ GeV. For $\tan\beta=20$ we give two sets of
curves, the upper one corresponds to the maximal value of the
neutrino Yukawa coupling $Y^\nu$. Here and in other plots, the
straight horizontal line corresponds to the current experimental
limit $B(\mu\rightarrow e\gamma)_{exp}<1.2\times10^{-11}$
\cite{pdg}.}\label{FIG. 3}
\end{figure}

In Figure \ref{FIG. 3}, we plot the branching ratio for the
process $\mu\rightarrow e\gamma$ including all the LFV effects
described earlier for Model 1 as a function of $M_{1/2}$ for
different values of $\tan\beta$ and $A_0$. We see that the
branching ratio is in the experimentally interesting range for
most of the parameter space. In this Figure for $\tan\beta=20$ we
also show the branching ratio when the Dirac neutrino Yukawa
coupling effects are maximized (denoted by ``large $Y^\nu$''). The
horizontal line corresponds to the current experimental limit.

\begin{figure}[!h]
\vspace*{-0.8truecm}
\leavevmode
\par
\begin{center}
\epsfig{file=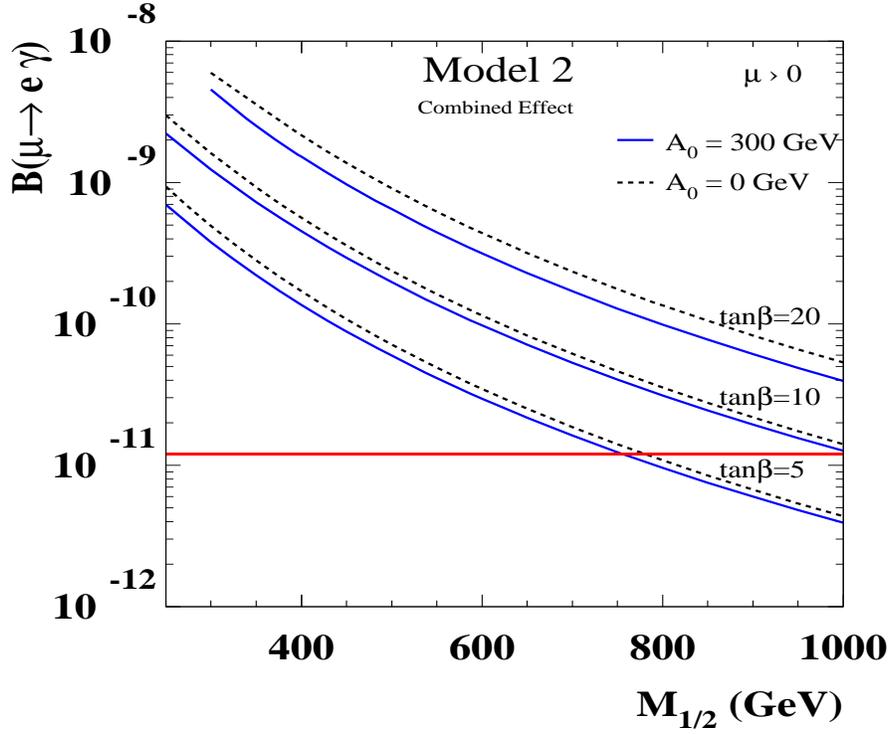,height=3.8in,width=4.6in}
\end{center}
\caption{\footnotesize Branching ratio for the process
$\mu\rightarrow e\gamma$ including all corrections for Model
2.}\label{FIG. 4}
\end{figure}

In Figure \ref{FIG. 4} the combined effect for $\mu\rightarrow
e\gamma$ is plotted for Model 2.

\begin{figure}[!h]
\vspace*{-0.5truecm}
 \leavevmode
\begin{center}
\epsfig{file=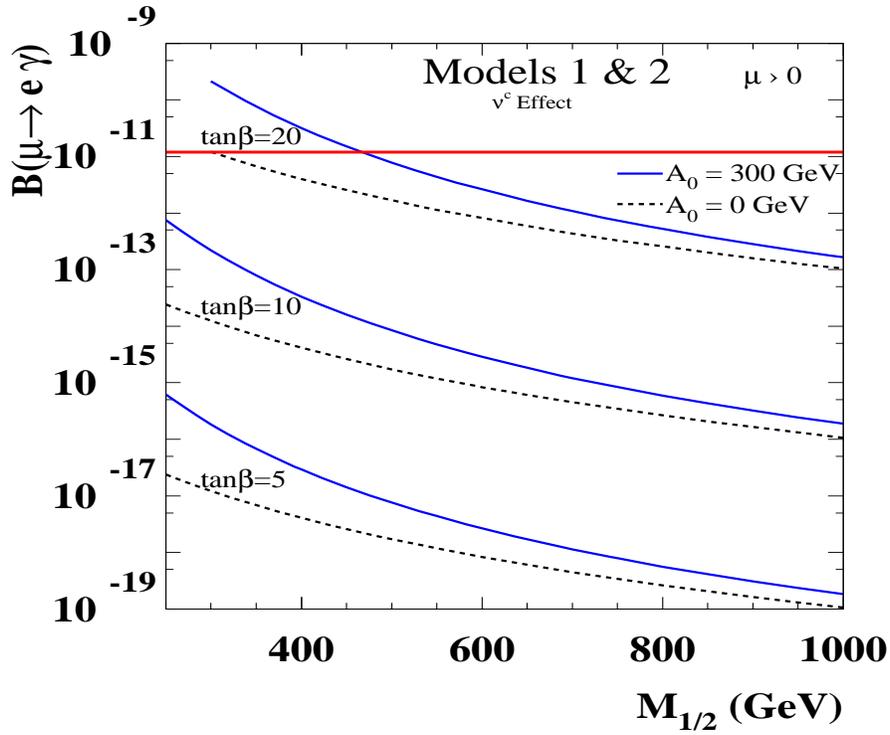,height=3.8in,width=4.6in}
\end{center}
\caption{\footnotesize Branching ratio for the process
$\mu\rightarrow e\gamma$ induced by only the right--handed
neutrino Yukawa coupling effects. This result holds for both
Models 1 and 2.}\label{FIG. 5}
\end{figure}

\begin{figure}[!h]
 \leavevmode
\vspace*{-0.8truecm}
\begin{center}
\epsfig{file=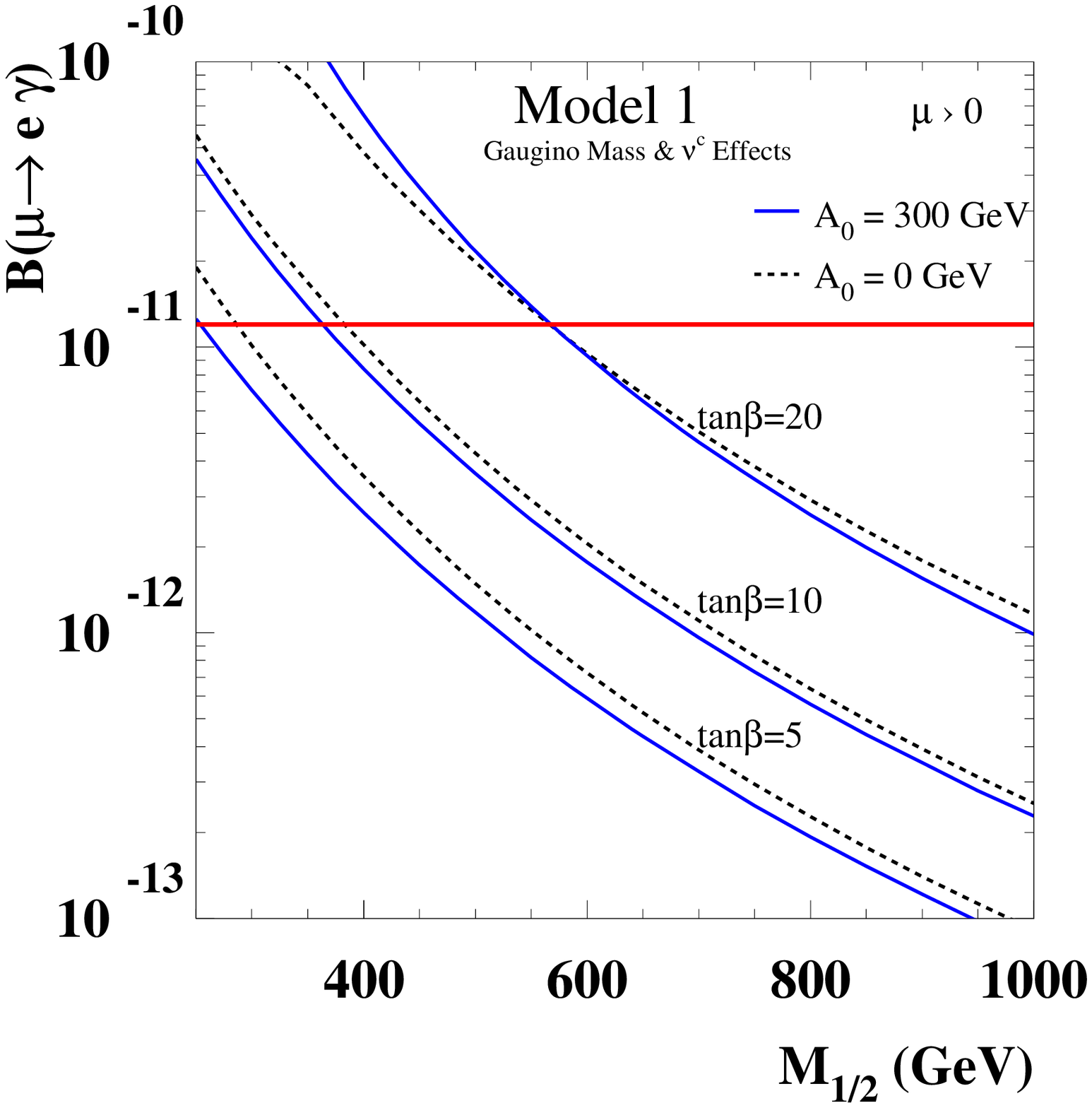,height=3.8in,width=4.6in}
\end{center}
\caption{\footnotesize Branching ratio for the process
$\mu\rightarrow e\gamma$ induced by the gaugino corrections (plus
$\nu^c$ effects) for Model 1. }\label{FIG. 6}
\end{figure}

\begin{figure}[!h]
\vspace*{-0.5truecm}
\begin{center}
\epsfig{file=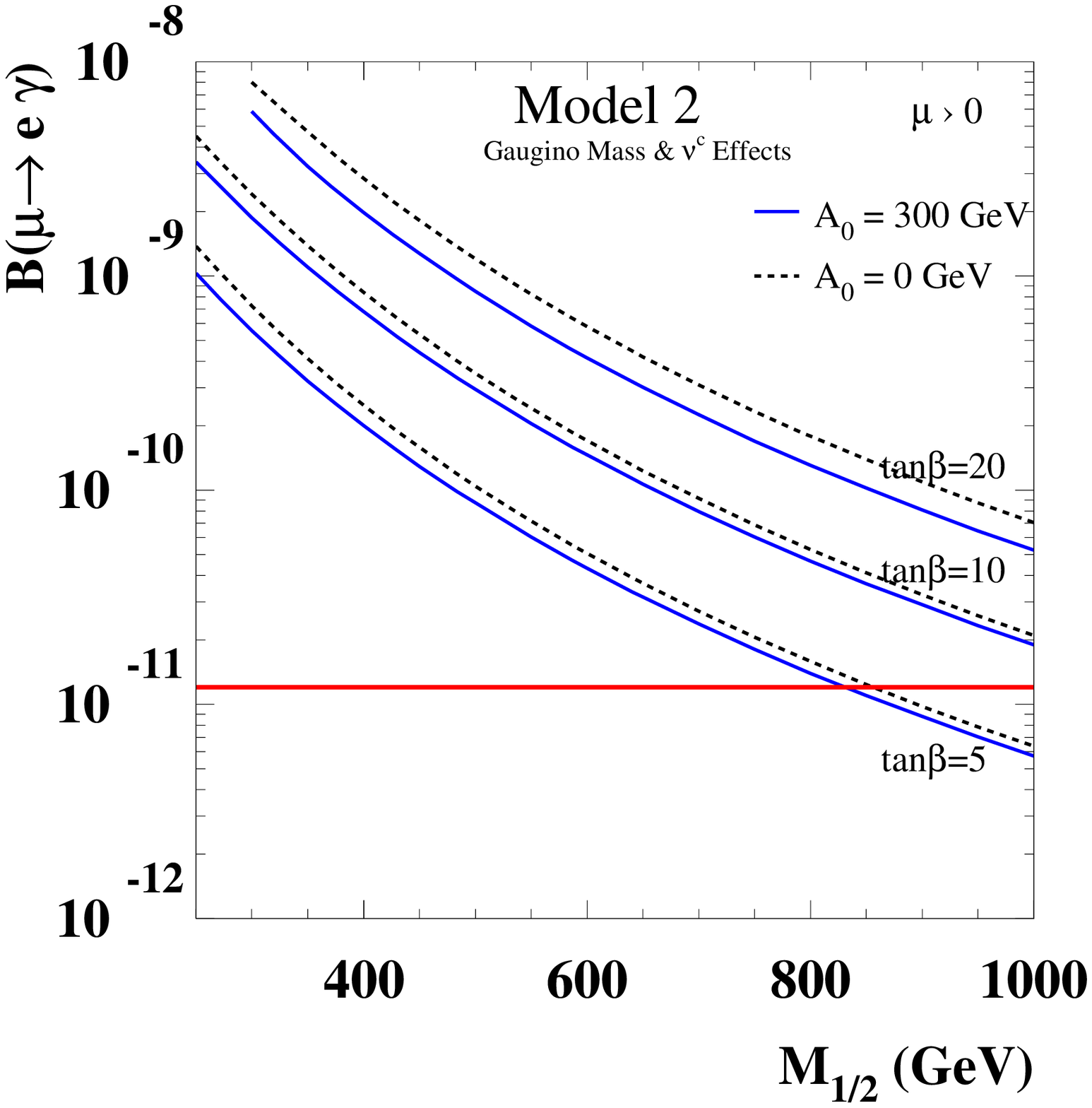,height=3.8in,width=4.6in}
\end{center}
\caption{\footnotesize Branching ratio for the process
$\mu\rightarrow e\gamma$ induced by the gaugino corrections (plus
$\nu^c$ effects) for Model 2. }\label{FIG. 7}
\end{figure}

In Figure \ref{FIG. 5} we plot $B\left(\mu\rightarrow
e\gamma\right)$ induced solely by the right--handed neutrino
Yukawa couplings. This result is identical for Models 1 and 2
since neutrino textures are the same for the two models. In Figure
\ref{FIG. 6} we plot the branching ratio induced by the
right--handed neutrino effects and the flavor gaugino effects for
Model 1. Figure \ref{FIG. 7} has the same plot for Model 2. In
Figure 8 (9) we plot $B\left(\mu\rightarrow e\gamma\right)$
induced by the trace term and the right--handed neutrino for Model
1 (2). Figure 10 (11) is a plot of the branching ratio including
the effects of $A$--terms and $\nu^c$ for Model 1 (2). Figures
\ref{FIG. 12} and \ref{FIG. 13} are the branching ratios for
$\tau\rightarrow \mu\gamma$ including all LFV effects for Model 1
and Model 2. Figure \ref{FIG. 14}, which is valid for both Models
1 and 2, has the branching ratios for $\tau\rightarrow \mu\gamma$
induced only by the $\nu^c$ Yukawa coupling effects.

From these figures we see that the decay $\mu\rightarrow e\gamma$
is within the reach of forthcoming experiments. Discovery of
$\tau\rightarrow \mu\gamma$ decay will strongly hint, within our
framework, an origin related to the right--handed neutrino Yukawa
couplings.

\newpage

\newpage

\newpage

\newpage

\begin{figure}[!h]
\vspace*{-0.8truecm}
\newpage
\begin{center}
\epsfig{file=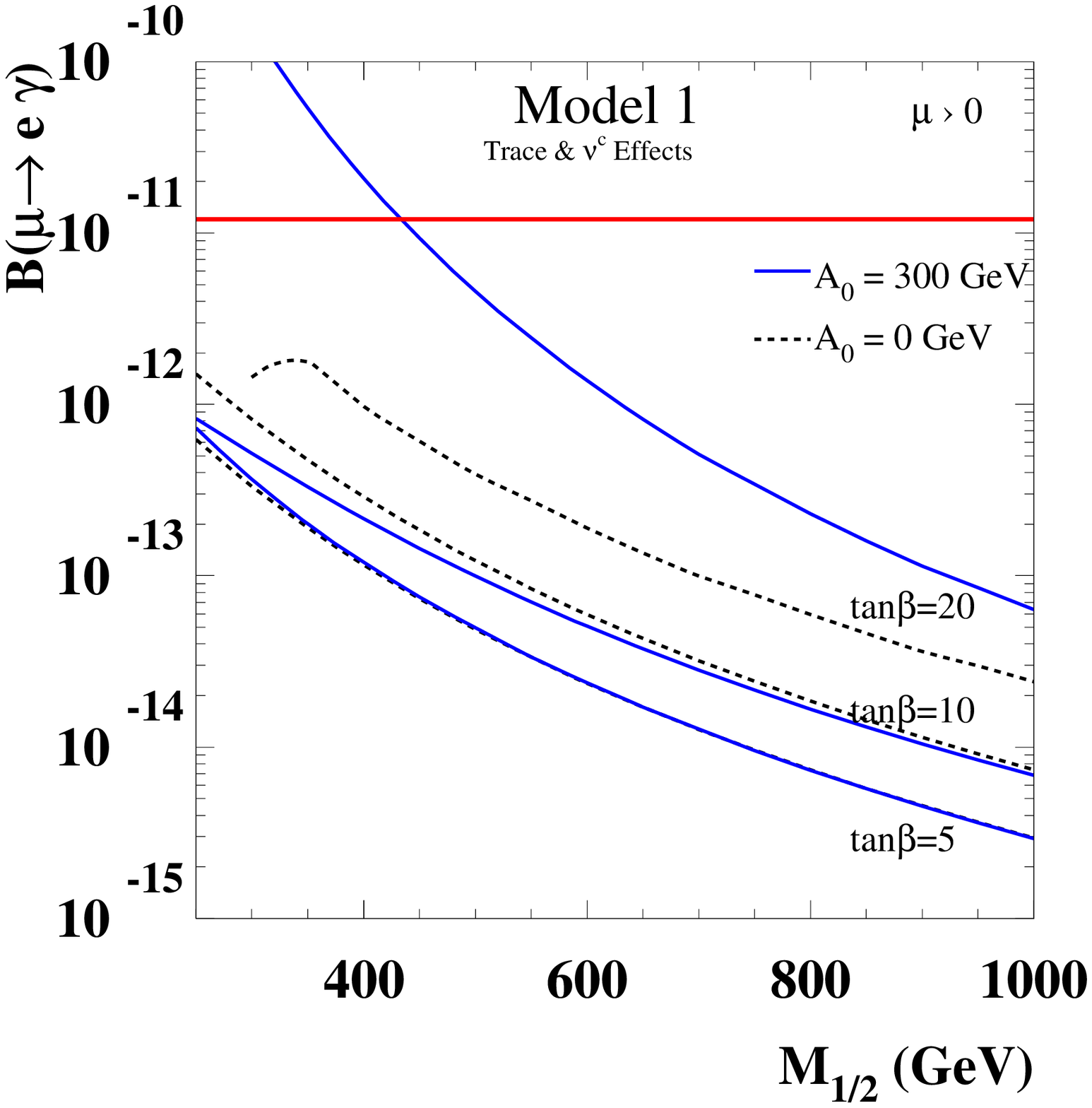,height=4.in,width=4.6in}
\end{center}
\caption{\footnotesize Branching ratio for the process
$\mu\rightarrow e\gamma$ induced by the trace correction (plus
$\nu^c$ effects) for Model 1.}\label{FIG. 8}
\end{figure}

\begin{figure}[!h]
\vspace*{-0.3truecm}
\begin{center}
\epsfig{file=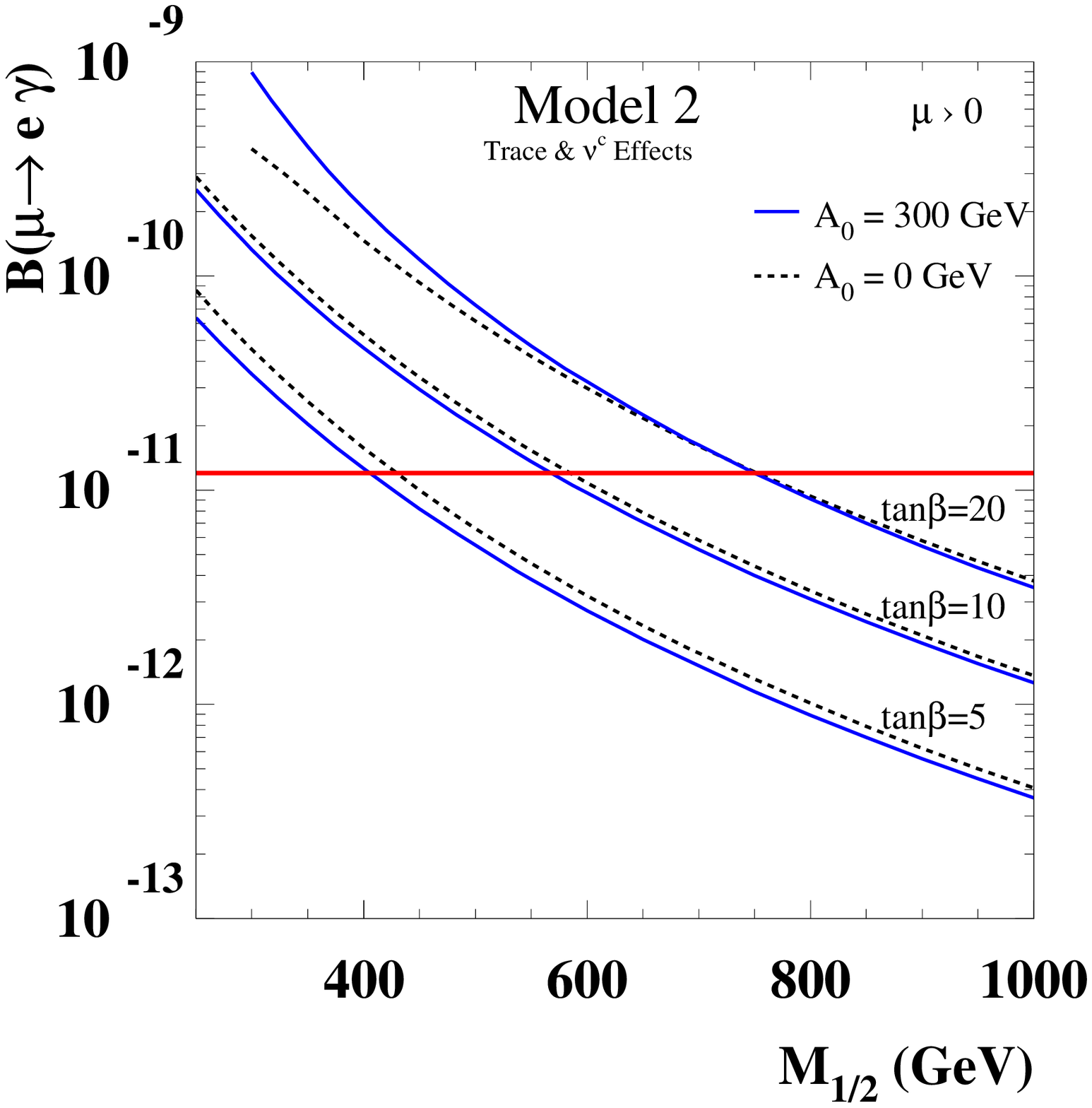,height=3.9in,width=4.6in}
\end{center}
\caption{\footnotesize Branching ratio for the process
$\mu\rightarrow e\gamma$ induced by the trace correction (plus
$\nu^c$ effects) for Model 2.}\label{FIG. 9}
\end{figure}

\newpage

\begin{figure}[!h]
\vspace*{-0.8truecm}
\begin{center}
\epsfig{file=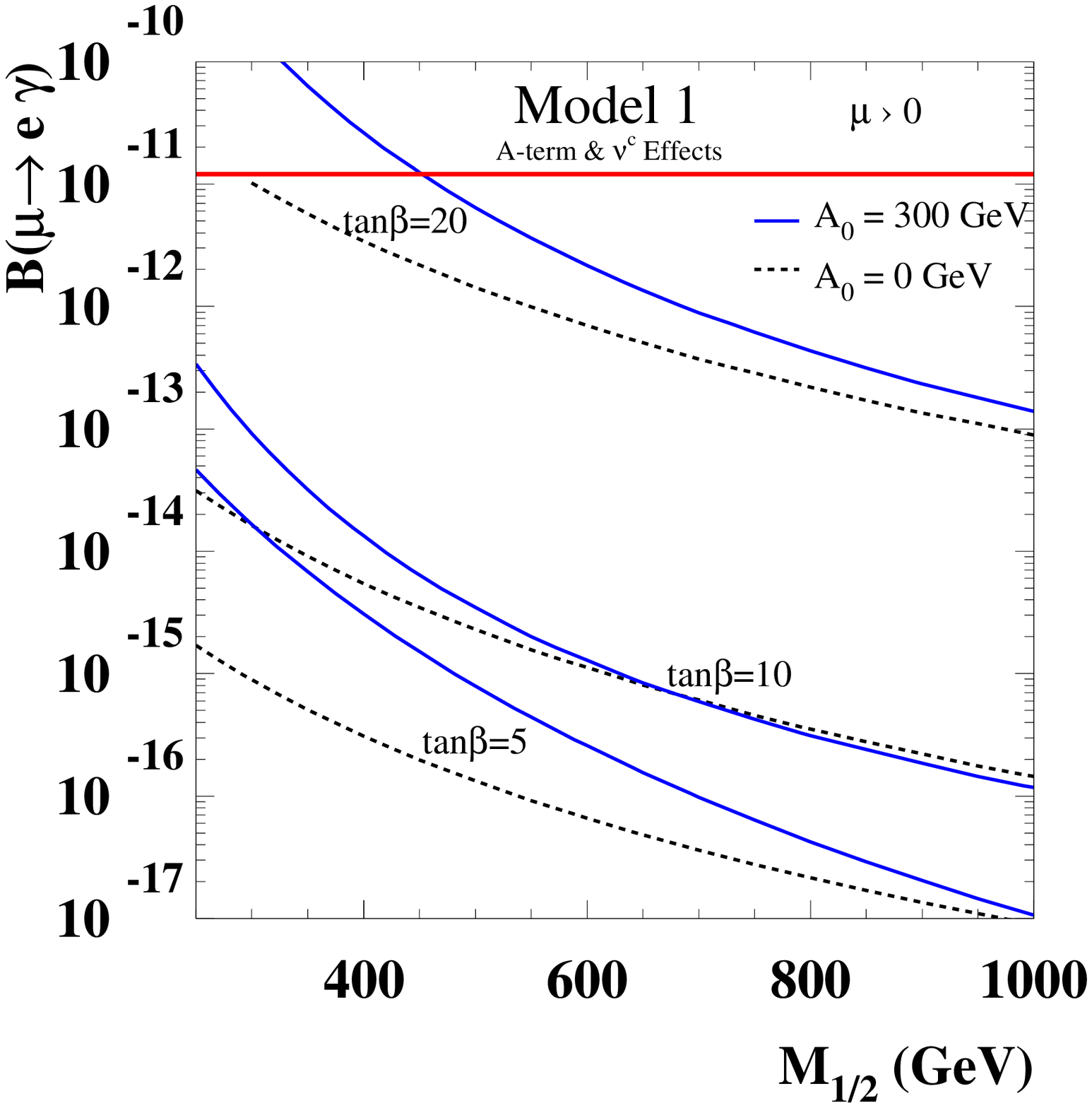,height=3.9in,width=4.6in}
\end{center}
\caption{\footnotesize Branching ratio for the process
$\mu\rightarrow e\gamma$ from the vertex corrections (plus $\nu^c$
effects) for Model 1.}\label{FIG. 10}
\end{figure}

\begin{figure}[!h]
\vspace*{-0.truecm}
\begin{center}
\epsfig{file=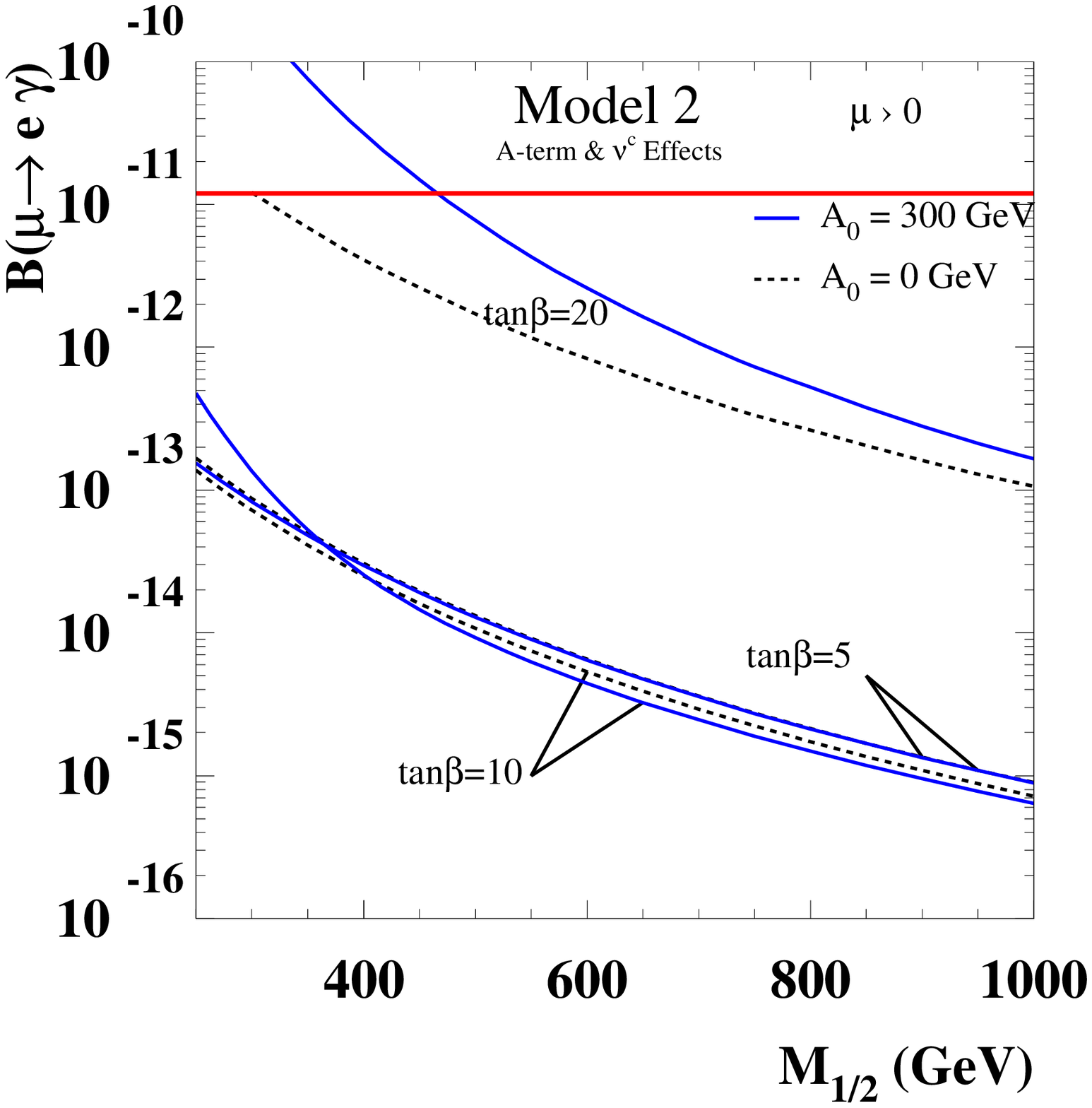,height=3.9in,width=4.6in}
\end{center}
\caption{\footnotesize Branching ratio for the process
$\mu\rightarrow e\gamma$ from the vertex corrections (plus $\nu^c$
effects) for Model 2.}\label{FIG. 11}
\end{figure}

\newpage

\begin{figure}[!h]
\vspace*{-0.8truecm}
\begin{center}
\epsfig{file=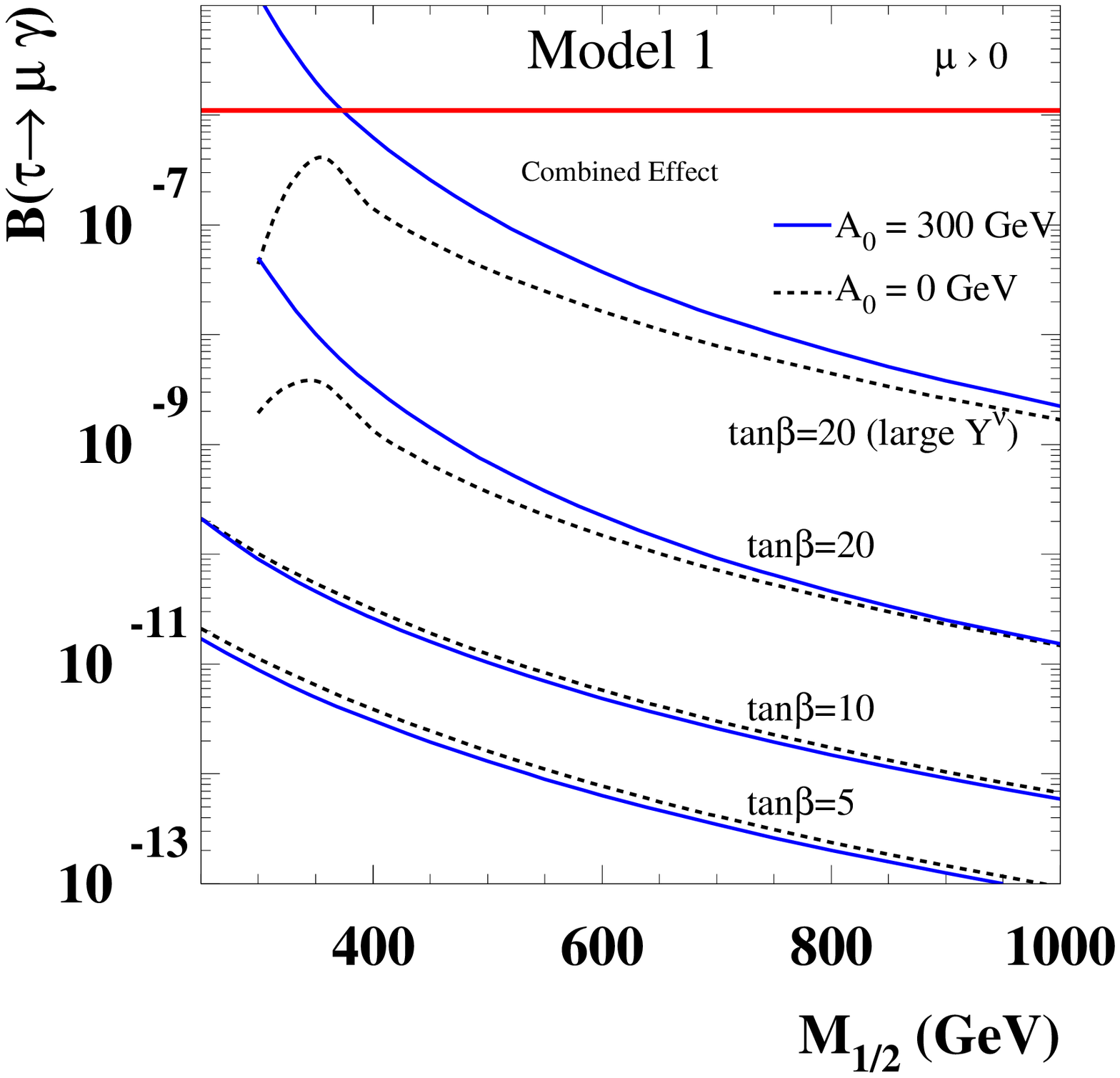,height=3.9in,width=4.6in}
\end{center}
\caption{\footnotesize Branching ratio for the process
$\tau\rightarrow\mu\gamma$ including all the effects for Model
1.}\label{FIG. 12}
\end{figure}

\begin{figure}[!h]
\vspace*{0.5truecm}
\begin{center}
\epsfig{file=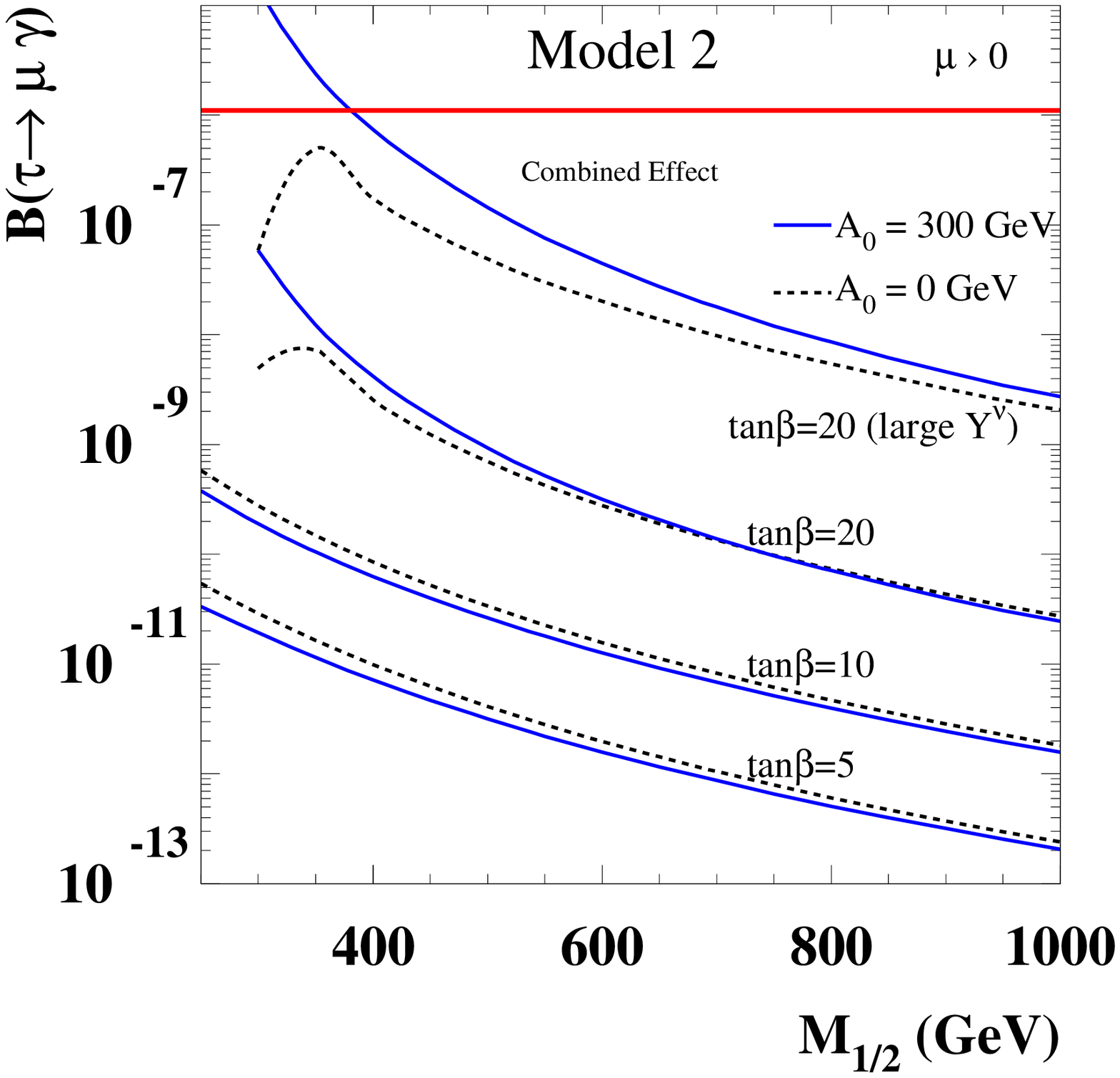,height=3.7in,width=4.6in}
\end{center}
\caption{\footnotesize Branching ratio for the process
$\tau\rightarrow\mu\gamma$ including all the effects for Model
2.}\label{FIG. 13}
\end{figure}

\newpage

\begin{figure}[!h]
\vspace*{-0.8truecm}
\begin{center} \epsfig{file=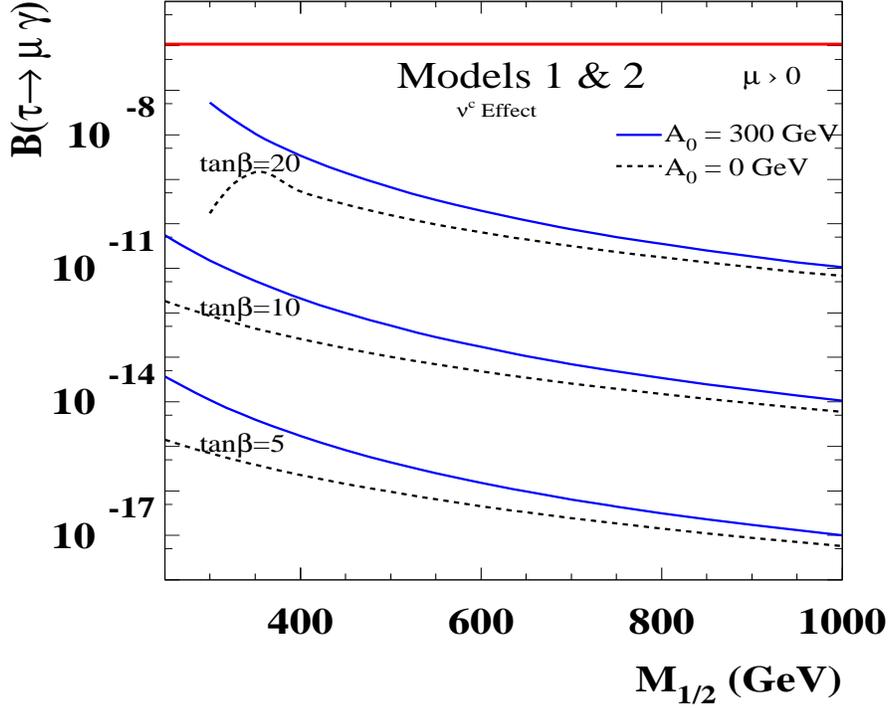,height=3.7in,width=4.6in}
\end{center}
\caption{\footnotesize Branching ratio for the process
$\tau\rightarrow\mu\gamma$ induced by only the right--handed
neutrino Yukawa coupling effects for Models 1 and 2.}\label{FIG.
14}
\end{figure}

\subsection{The D--term Splitting Problem}

Any model based on a gauged flavor symmetry has a potential
$D$--term splitting problem, which could give rise to large FCNC
processes even with a universal choice of soft scalar masses. Here
we quantify this problem in the class of anomalous $U(1)$ models.
We point out that this problem is not as serious as it might
naively appear, if the hierarchy $m_0=M_{1/2}/4.4$ needed for an
acceptable relic abundance of neutralino dark matter is assumed.

The $D_A$--term contribution to the scalar potential including
soft SUSY breaking mass for the flavon field $S$ is given by
\begin{eqnarray}\label{sclp}
V=\tilde{m}^2_s|S|^2+\frac{|q_s|^2g_F^2}{8}\left(\frac{\xi}{|q_s|}-|S|^2+\sum_iq_i|\tilde{f}_i|^2\right)^2
\end{eqnarray}
where $q_s$ and $q_{i}$ are the $U(1)_A$ flavor charges of the $S$
field and the MSSM fields $\tilde{f}_i$. Minimizing the potential
one finds the mass splitting among the MSSM sfermions to be
\begin{eqnarray}\label{softm1}
\tilde{m}_{f_i}^2-\tilde{m}_{f_j}^2=\frac{q_i-q_j}{|q_s|}\tilde{m}_s^2.
\end{eqnarray}
If a universal scalar mass is assumed even for $\tilde{m}^2_s$,
this splitting could be unacceptably large. However, if we choose
$m_0=M_{1/2}/4.4$, this may be in an acceptable range. The low
energy values for the slepton soft masses including the
$D_A$--term corrections to Eq. (\ref{SoftMapprx1}) are
\begin{eqnarray}\label{softm2}
\left(\tilde{m}_L^2\right)_{ii}=\frac{q^L_i}{|q_s|}\tilde{m}^2_s+m_0^2+0.52\,M_{1/2}^2\simeq\frac{q^L_i}{|q_s|}\tilde{m}^2_s+11.1\,m_0^2\,,\nn\\
\left(\tilde{m}_e^2\right)_{ii}=\frac{q^e_i}{|q_s|}\tilde{m}^2_s+m_0^2+0.15\,M_{1/2}^2\simeq\frac{q^e_i}{|q_s|}\tilde{m}^2_s+3.9\,m_0^2\,.
\end{eqnarray}
Since the $D$--terms only contribute to the diagonal slepton mass
splittings, any flavor violation arising from it must involve
fermionic mixing angles. We find
\begin{eqnarray}
\left(\delta^{RR}\right)^e_{12}&\simeq&\frac{\delta
q^e_{12}Y^e_{21}\tilde{m}_s^2}{Y^e_{22}
\left(\tilde{m}_e^2\right)_{22}}\,,\\
\left(\delta^{LL}\right)^e_{12}&\simeq&\frac{\delta
q^L_{12}Y^e_{12}\,\tilde{m}_s^2}{Y^e_{22}\left(\tilde{m}_L^2\right)_{22}}\,,
\end{eqnarray}
where $\left(\delta^{RR}\right)^e_{12}$ is the
$\tilde{e}_R$--$\tilde{\mu}_R$ mixing parameter in the
supersymmetric basis. Here $\delta
q^e_{12}=(q^e_1-q^e_2)/|q_s|=2-\alpha$ and $\delta
q^L_{12}=(q^L_1-q^L_2)/|q_s|=1$, with $\alpha=0 \,(1)$ for Model 1
(Model 2). For $\mu\rightarrow e \gamma$ this gives
\begin{eqnarray}
\left(\delta^{RR}\right)^e_{12}&\simeq&C_R\,\frac{(2-\alpha)\e^{1-\alpha}\left(\frac{\tilde{m}_s}
{m_0}\right)^2}{1+\frac{1}{1.95}\left(\frac{\tilde{m}_s}{m_0}\right)^2}\,,\\
\left(\delta^{LL}\right)^e_{12}&\simeq&C_L\,\frac{\left(\frac{\tilde{m}_s}{m_0}\right)^2}{
1+\frac{p}{11.1}\left(\frac{\tilde{m}_s}{m_0}\right)^2}\,,
\end{eqnarray}
where
\begin{eqnarray}\label{ClCr}
&&C_R=\frac{\e \,Y^e_{21}}{3.9\,Y^e_{22}}\simeq\left(\frac{1}{86.9},\,\frac{1}{46.9}\right),\nn\\
&&C_L=\frac{\e
\,Y^e_{12}}{11.1\,Y^e_{22}}\simeq\left(\frac{1}{75.6},\,\frac{1}{72.}\right)\,\,\,\mbox{for
(Model 1, Model 2)}.
\end{eqnarray}
From the experimental bound on $\mu\rightarrow e\gamma$ one
approximately has \cite{WellsTobe2}
\begin{eqnarray}
\left(\delta^{LL}\right)^e_{12}\left|_{exp}\approx\left(\delta^{RR}\right)^e_{12}\right|_{exp}
<\,1.2\times10^{-2}\frac{1}{\tan\beta}\left(\frac{M_{SUSY}}{500\,\mbox{GeV}}\right)^2.
\end{eqnarray}
For $\tan\beta=5$ and for $M_{SUSY}\simeq M_{1/2}$ (which is a
reasonable choice) we find
\begin{eqnarray}
\left(\delta^{LL,RR}\right)^e_{12}\left|_{exp}<0.6\times10^{-3}\,\left(1.0\times10^{-2}\right)\right.\,\,
\mbox{for}\,\,M_{1/2}=250\,\left(1000\right)\,\mbox{GeV}\,.
\end{eqnarray}
This gives the following constraint on $\tilde{m}_s/m_0$:
\begin{eqnarray}
\frac{\tilde{m}_s}{m_0}<0.2\,\left(0.9\right)\,\,
\mbox{for}\,\,M_{1/2}=250\,\left(1000\right)\,\mbox{GeV}\,
\end{eqnarray}
for Model 1 and
\begin{eqnarray}
\frac{\tilde{m}_s}{m_0}<0.17\,\left(0.7\right)\,\,
\mbox{for}\,\,M_{1/2}=250\,\left(1000\right)\,\mbox{GeV}\,
\end{eqnarray}
for Model 2. For low values of $M_{1/2}$ this gives a significant
constraint on $\tilde{m}_s$, while for large values of $M_{1/2}$
universality of all scalar masses may be maintained.

We wish to note that the $\tilde{m}_s^2$ appearing in Eq. (59) and
in the subsequent discussions is the soft mass--squared of the
scalar flavon field $S$ evaluated at the $U(1)_A$ breaking scale
$M_F$. Since there are singlet fields $X_k$ in the theory needed
for anomaly cancellation which are active between the $U(1)_A$
scale and $M_{st}$ (Cf: Eq. (9)), the $S$ field can have
renormalizable Yukawa couplings of the type $Y_k X_k \bar{X_k} S$
where $\bar{X_k}$ are fields neutral under $U(1)_A$. We have
examined the evolution of $\tilde{m}_s^2$ between $M_F$ and
$M_{st}$ in the presence of such Yukawa couplings and found that
even with the limits on $\tilde{m}_s$ given in Eqs. (68) and (69),
$\tilde{m}_s^2 = m_0^2$ can be attained at the string scale for a
range of Yukawa couplings $Y_k$.

The RGE running of $\tilde{m}_s^2$ between the string scale and
the $U(1)_A$ symmetry breaking scale contains the same trace term
and the same gaugino contributions that we considered for the MSSM
sfermions in Eqs. (\ref{deltams1a}) and (\ref{deltams1g}) (see
also terms proportional to $\sigma$ in Eqs. (36)-(41))
\begin{eqnarray}\label{dms1}
&&\delta\tilde{m}_s^2\simeq -q_s|q_s|g_F^2
\left(3\,m_0^2\,\sum_{i=1,3}\left(n^u_{ii}+n^d_{ii}\right)\nn\right.\\&&\,\,\,\,\,\,\,
\left.+m_0^2\,\sum_{i=1,3}\left(n^e_{ii}+n^\nu_{ii}\right) +n^X
m_0^2-\tilde{m}_s^2\right)\frac{{\rm
ln}\left(M_{st}/M_F\right)}{8\pi^2}\nn\\
&&\,\,\,\,\,\,\,+\left(q_s\,g_F\right)^2\left(M_{\lambda_F}\right)^2\frac{{\rm
ln}\left(M_{st}/M_F\right)}{2\pi^2}\,. \end{eqnarray}
Interestingly,  this contribution from the $D$-term evolution
cancels the ``trace contributions" in the sfermion masses (the
terms proportional to $\sigma$ in Eqs. (36)-(41))
\cite{kobayashi}.

\section{Conclusion}

In this paper we have studied lepton flavor violation induced by a
flavor--dependent anomalous $U(1)$ gauge symmetry of string origin
in a class of models which addresses the fermion mass hierarchy
problem via the Froggatt--Nielsen mechanism.  We have derived a
general set of renormalization group equations for the evolution
of soft SUSY breaking parameters in the presence of higher
dimensional operators.  These results should be applicable to a
large class of fermion mass models.  We have shown that the
$U(1)_A$ sector induces significant flavor violation in the SUSY
breaking parameters during the RGE evolution from the string scale
to the flavor symmetry breaking scale, even though this momentum
range is very short.  We have identified several sources of flavor
violation:  the $U(1)_A$ gaugino contribution to the scalar masses
which is flavor dependent, a contribution proportional to the
trace of $U(1)_A$ charge which is also flavor dependent,
non--proportional $A$--terms arising from the $U(1)_A$ gaugino
vertex correction diagrams, and the $U(1)_A$ $D$--term. In
addition, there are flavor violating effects in the charged lepton
sector arising from the right--handed neutrino Yukawa couplings,
which have also been included in our numerical analysis. The
resulting flavor violation in the leptonic decays $\mu \rightarrow
e \gamma$ and $\tau \rightarrow \mu \gamma$ are found to be in the
experimentally interesting range.

Adopting the minimal supergravity scenario for SUSY breaking, and
choosing parameters such that the needed relic abundance of
neutralino dark matter is realized, we have presented results for
the branching ratios $B(\mu \rightarrow e \gamma)$ and $B(\tau
\rightarrow \mu \gamma)$ in two specific models of fermion masses.
Figures 3 and 4 are our main results for the two models for $B(\mu
\rightarrow e \gamma)$, while Figures 13 and 14 are our results
for $B(\tau \rightarrow \mu \gamma)$.  The former should be
accessible to forthcoming experiments, while the latter is also in
the observable range. Although we focused on two specific fermion
mass textures these effects should be significant in a large class
of models.

\section*{Acknowledgments}

We thank C. Macesanu for helpful discussions. This work is
supported in part by DOE Grant \# DE-FG03-98ER-41076, an award
from the Research Corporation and by DOE Grant \#
DE-FG02-01ER-45684.

\end{document}